\documentclass[aps,twocolumn]{revtex4-1}

\usepackage{amsmath}
\usepackage{amsfonts}
\usepackage{amssymb}
\usepackage{color}
\usepackage{graphicx}

\newcommand{\cmt}{\bar t}

\begin{document}

\title{A thermally driven spin-transfer-torque system far from equilibrium: enhancement of the thermoelectric current via pumping current}
%\title{In thermally driven spin-transfer-torque systems the pumping current can enhance the thermoelectric current} 
%alternative: Thermally driven spin torque systems far from equilibrium \\ --- enhancement of thermoelectric effect via pumping current

\author{Tim \surname{Ludwig}$^1$, Igor S. \surname{Burmistrov}$^{2,3,1,4}$, Yuval \surname{Gefen}$^5$, 
Alexander \surname{Shnirman}$^{1,4}$}
\affiliation{$^1$Institut f\"ur Theorie der Kondensierten Materie, Karlsruhe Institute of Technology, 76128 Karlsruhe, Germany}
\affiliation{$^2$L.D. Landau Institute for Theoretical Physics RAS, Kosygina street 2, 119334 Moscow, Russia}
\affiliation{$^3$Laboratory for Condensed Matter Physics, National Research University Higher School of Economics, 101000  Moscow, Russia}
\affiliation{$^4$Institut für Nanotechnologie, Karlsruhe Institute of Technology, 76021 Karlsruhe, Germany}
\affiliation{$^5$Department of Condensed Matter Physics, Weizmann Institute of Science, 76100 Rehovot, Israel}

\begin{abstract}
We consider a small itinerant ferromagnet exposed to an external magnetic field and strongly driven by a thermally induced spin current. For this model, we derive the quasi-classical equations of motion for the magnetization where the effects of a dynamical non-equilibrium distribution function are taken into account self-consistently. We obtain the Landau-Lifshitz-Gilbert equation supplemented by a spin-transfer torque term of Slonczewski form. We identify a regime of persistent precessions in which we find an enhancement of the thermoelectric current by the pumping current. 
\end{abstract}

\maketitle

\section{Introduction}
The field of spintronics can be very roughly summarized as dealing with the manipulation of magnets and spin-currents by use of charge currents and vice versa \cite{wolf2001spintronics, vzutic2004spintronics, tserkovnyak2005nonlocal}. Inclusion of thermal transport effects into spintronics gives rise to the field of spin-caloritronics which is not only of fundamental interest but also of technical relevance: an efficient conversion of heat flow into a more useful form of energy would be of particular interest for the technical reuse of otherwise wasted heat \cite{bauer2012spin,boona2014spin}. Spin-caloritronic effects are roughly classified into~\footnote{There is also the class of relativistic effects, which, however, do not play any role in this work.} single particle effects, like standard Seebeck and Peltier effect but with spin-dependent density of states, and collective effects (magnons) \cite{bauer2012spin, boona2014spin}.

Spincaloritronic effects in magnetic tunnel-junctions are often considered in terms of single particle effects, see for example refs. \cite{lin2012giant, walter2011seebeck}. Recently, it was shown that collective effects can become very important in the description of magnetic tunnel-junctions \cite{PhysRevB.96.094429}. In those works, the magnetic tunnel-junctions are described as two magnetic leads tunnel-coupled to each other (F$|$I$|$F). A non-equilibrium situation is generated by assuming a different temperature in each magnet. This is reasonable for two magnets that are large enough for an equilibrium distribution of elementary excitations to develop in the vicinity of the tunneling contact, even under the influence of the driving force. In contrast, we consider a small itinerant ferromagnet placed in between an itinerant ferromagnetic lead and a normal metal (F$|$I$|$F$|$I$|$N), Fig. \ref{fig: system}. 
For mesoscopic systems, it is important to include non-equilibrium effects in the distribution function, 
when considering a small system placed between two leads. In spin-caloritronics, these non-equilibrium
effects have been addressed recently in ref. \cite{PhysRevB.98.014406} . The central theme of our work is the interplay of those
non-equilibrium effects with the dynamics of the magnetization. To our knowledge, this has not yet been studied for spin-caloritronic systems.
 
Heading into this new direction of strong non-equilibrium effects in spin-caloritronic systems, we keep the magnetic part of the model quite simple (e.g. no internal magnetic anisotropy). We expect the non-equilibrium picture developed here to be of more universal validity.

We describe the small itinerant ferromagnet with dynamical magnetization by the universal Hamiltonian of ref. \cite{PhysRevB.62.14886}. Instead of a proper (internal) magnetic anisotropy, we consider an external magnetic field only. We assume the system to be deep in the Stoner-regime with a large magnetization (respectively spin) and we use the macrospin approximation, i.e., only the Kittel mode is considered. The large spin renders the dynamics of the angular part of the magnetization quasi-classical. 
The magnetization of the ferromagnetic lead is fixed and parallel to the external magnetic field. We assume many channels in the leads with spin-independent tunnel-coupling to the small magnet, so that the dimensionless conductance of each junction is large and the Coulomb-blockade is exponentially supressed. This allows for a quasi-classical description of the dynamics of the magnetization length and the electrical potential of the small itinerant ferromagnet.
A non-equilibrium situation is generated by a temperature difference in the leads, and we disregard internal relaxation mechanisms, which puts our model in the regime opposite to refs. \cite{lin2012giant, walter2011seebeck, PhysRevB.96.094429}. 

While the model as a whole may be too naive for real spin-transfer-torque systems, it allows us to focus on the interplay of magnetization dynamics and the dynamic non-equilibrium distribution function in the small itinerant magnet.
Extending the ideas of refs. \cite{PhysRevLett.114.176806, PhysRevB.95.075425}, we derive an effective quasi-classical action of a generalized Ambegaokar-Eckern-Sch\"on type \cite{AES_PRL, AES_PRB} ($U(1)\otimes U(1)\otimes SU(2)$) for the electrical potential and the magnetization jointly. 
For the quasi-classical angular dynamics of the magnetization we obtain the Landau-Lifshitz-Gilbert equation including a spin-transfer torque term of the Slonczewski form \cite{slonczewski1996current}. We also determine the stationary charge current flowing through the system. We share the conclusion of ref. \cite{PhysRevB.96.094429}, namely, that collective effects are important in magnetic tunnel-junctions. In particular, we identify single-particle effects and collective contributions to be important for both, the spin-tranfer-torque and the charge current. More explicitly, in the regime of persistent precession the pumped current (a collective effect) can enhance the thermoelectric effect.

Finally, we note that, apart from the nature of the driving bias (thermal vs. electrical), the system discussed here is identical to that of ref. \cite{PhysRevB.95.075425}. For the sake of convenience, we repeat here the essential parts of the derivation. However, a regime of persistent precession remains which makes it necessary to go beyond ref. \cite{PhysRevB.95.075425}, which we extend to allow for a simplified treatment of the angular dynamics of the magnetization.

This article is organized as follows: In section \ref{sec: the system} we introduce the Hamiltonian of the system and discuss the distribution functions of the leads. Making use of gauge transformations, we formally derive an effective quasi-classical action in section \ref{sec: effective action}. In section \ref{sec: determine GF} we determine the classical Green's function, which is used in section \ref{sec: EOM} to obtain the quasi-classical equations of motion. Finally, the charge current flowing through the system is determined in section \ref{sec: currents}, where we also discuss the enhancement of the thermoelectric effect.

\section{The System \label{sec: the system}}
\begin{figure}
\begin{center}
\includegraphics[width=0.45\textwidth]{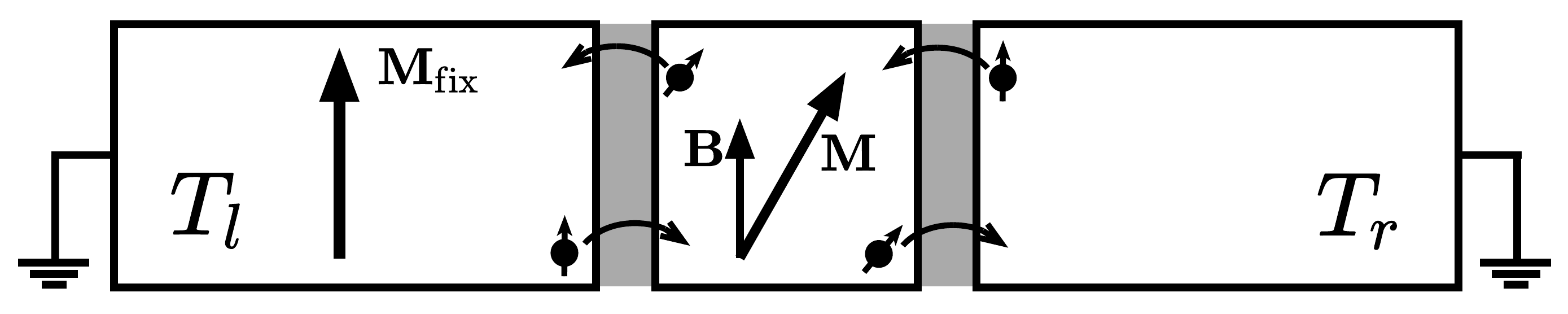}
\end{center}
\caption{A schematic view of the system: A small (0-dimensional) itinerant ferromagnet is placed in an external magnetic field and tunnel-coupled to two leads. One lead is magnetic with a fixed direction of magnetization (left), while the other lead is a normal metal (right). The system can be driven out of equilibrium by a temperature difference between the leads.}\label{fig: system}
\end{figure}
We consider an itinerant ferromagnetic quantum dot which is exposed to an external magnetic field and tunnel-coupled to two leads, see Fig. \ref{fig: system}. The left lead is an itinerant-ferromagnet itself but with a fixed magnetization. The right lead is a normal metal. The system can be driven out of equilibrium by a temperature difference between the leads. The Hamiltonian of the full system is
\begin{equation}
H= H_\mathrm{dot} + H_l + H_r + H_\mathrm{tun.}\ .
\end{equation}

To describe the ferromagnetic quantum dot, we use the universal Hamiltonian \cite{PhysRevB.62.14886}, but disregard the interaction in the Cooper channel:
\begin{equation}
H_\mathrm{dot} = H_0 - J \mathbf S^2+ E_c (N-N_0)^2 - \mathbf B \mathbf S\ .
\end{equation}
The non-interacting part is $H_0=\sum_{\alpha \sigma} \epsilon_\alpha\, a_{\alpha \sigma}^\dagger a_{\alpha \sigma}^{\phantom\dagger}$, with $\alpha$ denoting single-particle states on the dot. The exchange interaction $- J \mathbf S^2$, with exchange constant $J$ and the total spin operator $\mathbf{S} = \frac{1}{2} \sum_{\alpha, \sigma_1, \sigma_2} a_{\alpha \sigma_1}^\dagger \boldsymbol \sigma_{\sigma_1 \sigma_2} a_{\alpha \sigma_2}^{\phantom\dagger}$, tends to align electron spins on the dot. The charging interaction, which accounts for repulsion of charges on the dot, is given by $+ E_c (N-N_0)^2$ with $E_c=\frac{1}{2C}$ and $C$ is the capacity, $N_0$ represents the positive background charges, and the total number operator is given by $N= \sum_{\alpha \sigma} a^\dagger_{\alpha \sigma} a^{\phantom \dagger}_{\alpha \sigma}$.
The coupling to the external magnetic field is described by the Zeeman-energy of the total spin $- \mathbf B \mathbf S$ and we choose the external magnetic field to be along the $z$-direction, i.e. $\mathbf B = (0,0,B)$.

The leads are described as non-interacting systems. The fixed magnetization of the left lead\footnote{For physical units, we have to replace $\mathbf B = \rightarrow \lambda \mathbf{B}$ and $M_\mathrm{fix} \rightarrow \lambda M_\mathrm{fix}$, where $\lambda = \frac{\hbar g e}{2 m c}$ with $g \approx 2$ and $e<0$. Similarly, for the Hubbard-Stratonovich fields introduced below $V_d \rightarrow e V_d$, $\mathbf B_\mathrm{exc} \rightarrow \lambda \mathbf B_\mathrm{exc}$, and $\mathbf M \rightarrow \lambda \mathbf M$.}, which is assumed to be parallel to the external magnetic field, is taken into account as a spin-dependent background-potential for electrons,
\begin{equation}
H_l= \sum_{n=1}^{N_l} \sum_\sigma \int \frac{dk}{2 \pi} \left(\epsilon_{nk} - \frac{M_\mathrm{fix}}{2} \sigma \right) c_{nk, \sigma}^\dagger c_{nk, \sigma}^{\phantom \dagger}\ ,
\end{equation}
where $- \frac{M_\mathrm{fix}}{2} \sigma$ accounts for the different energy of electrons with spin up versus spin down and $n= {1, ... , N_l}$ counts the channels for the left lead and $k$ denotes the momentum. The nonmagnetic right lead is described by,
\begin{equation}
H_r= \sum_{n=N_l+1}^{N_l+N_r} \sum_\sigma \int \frac{dk}{2 \pi}\ \epsilon_{nk}\  c_{nk, \sigma}^\dagger c_{nk, \sigma}^{\phantom \dagger}\ .
\end{equation}
Here $n= {N_l+1, ... , N_l+N_r}$ counts the channels for the right lead and $k$ denotes the momentum again. 

The tunneling between the dot and the leads is described by,
\begin{equation}
H_\mathrm{tun.} = \sum_{n=1}^{N_l+N_r} \sum_{\alpha \sigma} \int \frac{dk}{2\pi}\ t_{\alpha n}\ a^\dagger_{\alpha \sigma} c_{nk,\sigma} + h.c.\ ,
\end{equation}
where the tunneling amplitudes $t_{\alpha n}$ will include some randomness, since we have chosen to diagonalize the non-interacting part of the dot Hamiltonian $H_0$. 

The system is not yet fully specified. In addition to the Hamiltonian, we also have to know the distribution functions. We fix the distribution function of each lead to be a Fermi-distribution. For both, we choose the same electrochemical potential $\mu$, but allow for different temperatures $T_{l/r}$, i.e. $n_{l/r}(\epsilon) = 1/\left(e^{(\epsilon-\mu)/T_{l/r}}+1\right)$. In principle, we could also specify the initial distribution function of the dot. However, after a short time (of the same order as the life-time of electrons in the dot), the information about this initial distribution will be lost \cite{PhysRevB.82.155317}. Afterwards, the distribution function of the dot will be enslaved to both the distribution functions of the leads and the dynamics of magnetization and electrical potential on the dot\footnote{Thereby, we take a view complementary to the standard view of kinetic equation approaches, where the distribution function is the fundamental object and (the dynamics of) the magnetization would be determined from (the dynamics of) the distribution function. In turn, one could say that the magnetization is enslaved to the distribution function.}.
Since we are not interested in the initial transient effects, there is no need to specify the initial dot's distribution function. However, the enslaved but dynamic distribution function is crucial for the dynamics and will be determined below.

\section{The effective action \label{sec: effective action}}
We are dealing with a non-equilibrium situation and therefore the Keldysh formalism is employed \cite{KamenevBook,doi:10.1080/00018730902850504}. We use its path integral version. The Keldysh generating function is given by
\begin{equation}
\mathcal Z=\int D[\bar\Psi,\Psi] e^{i \mathcal S\left[\bar\Psi,\Psi\right]}\ ,
\end{equation}
where $\Psi,\bar\Psi$ denote fermionic fields. The action is given by
\begin{equation}
i \mathcal S\left[\bar\Psi,\Psi\right]= i \oint_K dt\, [\bar\Psi\, i\partial_t\, \Psi -H(\bar \Psi,\Psi)]\ ,
\end{equation}
where the integral is over the Keldysh contour\footnote{The Keldysh contour is chosen to go from $-T_K$ to $+T_K$ on the upper contour (+) and backwards on the lower contour (-).}. 

\subsection{Integrating out the leads}
The fermionic fields of the leads enter only up to quadratic order. Thus, the leads can be integrated out and we obtain,
\begin{equation}
i \mathcal S\left[\bar\Psi,\Psi\right]= i \oint_K dt\, [\bar\Psi\, (i\partial_t - \Sigma)\, \Psi -H_\mathrm{dot}(\bar \Psi,\Psi)]\ , \label{eq: Keldysh action}
\end{equation}
where $\Sigma = \Sigma_l + \Sigma_r$ is the self-energy related to the tunneling between the dot and the leads. The self-energies for the leads are given by $\Sigma_l = t_l G_l t^\dagger_l$ and $\Sigma_r = t_r G_r t^\dagger_r$; the lead Green's functions $G_{l/r}$ are defined by $G_{l/r}^{-1}= i \partial_t - H_{l/r}$. The tunneling matrix $t_l$ consists of elements $t_{\alpha n}$ with $n={1, ..., N_l}$ and similarly $t_r$ consists of elements $t_{\alpha n}$ with $n={N_l+1, ..., N_l+N_r}$.

We assume a large number of weakly and randomly coupled transport channels. Then, the tunneling can be approximately described by just three tunneling rates: $\Gamma_l^\uparrow, \Gamma_l^\downarrow$ for the spin-dependent coupling to the left lead and  $\Gamma_r$ for the coupling to the right lead \cite{PhysRevB.95.075425}. The tunneling rates are determined by the averaged tunneling amplitudes and the spin resolved densities of states at the electrochemical potential of the leads.

The effect of tunneling between leads and dot is twofold. First, it determines the life-time of the states of the dot. 
Second, the leads provide a heat and particle bath for the dot. The self-energy should, thus, carry information about the level-broadening as well as the respective electron distributions in the leads. Indeed, information about level-broadening is contained in the retarded and advanced part $\Sigma_{\sigma}^{R/A}(\omega)= \mp i (\Gamma_l^\sigma+\Gamma_r)$, whereas the Keldysh part carries the information about the distribution functions of the leads $\Sigma_{\sigma}^K(\omega) = - 2i (\Gamma_l^\sigma F_l(\omega) + \Gamma_r F_r(\omega))$, where $F_{l/r}(\omega)=1-2\, n_{l/r}(\omega)$. We emphasize that the distribution function of the dot does not appear explicitly in eq. \eqref{eq: Keldysh action}. It is enslaved to the distribution functions of the leads in combination with the magnetic dynamics of the dot.

\subsection{Decoupling of the interactions}
We decouple the interactions by performing a Hubbard-Stratonovich (HS) transformation. For the exchange interaction, we use,
\begin{equation}
e^{i J \oint_K dt\, \mathbf S^2} = \int D\mathbf B_\mathrm{exc}\ e^{-i \oint_K dt\, \left( \frac{\mathbf B_\mathrm{exc}^2}{4J} - \mathbf B_\mathrm{exc} \mathbf S\right)}\ ,
\end{equation}
and for the charging interaction, we use,
\begin{equation}
e^{-i E_c \oint_K dt\, \left(N-N_0\right)^2} = \int D V_d\ e^{i \oint_K dt\, \left(\frac{V_d^2}{4E_c} - V_d (N-N_0) \right)}\ ,
\end{equation}
which make the action quadratic in fermionic fields. Then, we can integrate out the fermions and, after reexponentiation, we obtain,
\begin{equation}
i \mathcal S= \mathrm{tr}\, \mathrm{ln} \left[ G_{0}^{-1} +\mathbf M \frac{\boldsymbol \sigma}{2} -V_d -\Sigma \right] + i \mathcal{S}_\mathrm{HS}\ ,  \label{eq: exact effective action}
\end{equation}
with
\begin{equation}
i \mathcal{S}_\mathrm{HS} = - i \oint_K\!\! dt \frac{(\mathbf M - \mathbf B)^2}{4J}+i \oint_K\!\! dt \left[\frac{V_d^2}{4E_c} +V_d N_0 \right]\ , \label{eq: original HS}
\end{equation}
and we defined $G_{0}^{-1} = i \partial_t -H_0$ and $\mathbf M = \mathbf B + \mathbf B_\mathrm{exc}$, to which we refer as the magnetization\footnote{$\mathbf B_\mathrm{exc}$ is proportional to the true magnetization but in the ferromagnetic case we have $|\mathbf B_\mathrm{exc}| \gg |\mathbf B |$ and therefore $\mathbf M \approx \mathbf B_\mathrm{exc}$.}. 

\subsection{The rotating frame}
The time-dependence of $\mathbf M$ in the $\mathrm{tr}\, \mathrm{ln}[...]$ renders the action in eq. \eqref{eq: exact effective action} quite non-trivial. To deal with this, we perform a transition into a rotating frame, in which 
$\mathbf M$ is at all times directed along the $z$-axis. This is the same SU(2)-gauge transformation, as in refs. \cite{PhysRevLett.114.176806, PhysRevB.95.075425}. For that purpose, we separate the magnetization $\mathbf M = M\, \mathbf m$ into its length $M = |\mathbf M|$ and its direction $\mathbf m$. Then, we introduce the spin-rotation matrix $R$, such that the magnetization is rotated onto the $z$-axis, i.e. $R^\dagger\mathbf{m} \boldsymbol \sigma R= \sigma_z$. Due to the time dependence of the direction $\mathbf m$ of the magnetization, the rotations $R$ will also depend on time. Therefore, performing the rotation comes on the cost of generating a new term $Q=-i R^\dagger \dot R$ due to the time derivative in $G_{0}^{-1}$. For the action we obtain,
\begin{equation}
 i \mathcal S=\mathrm{tr}\, \mathrm{ln} \left[ G_{0}^{-1} + M \frac{\sigma_z}{2} - V_d - R^\dagger \Sigma R - Q \right] + i \mathcal{S}_\mathrm{HS}\, . \label{eq: SU2 rotated action}
\end{equation}
To proceed, we choose the Euler angle representation,
\begin{equation}
R=e^{-i \frac{\phi}{2} \sigma_z} e^{-i \frac{\theta}{2} \sigma_y} e^{i \frac{\phi - \chi}{2} \sigma_z}\ , \label{eq: R}
\end{equation}
where $\chi$ is a gauge freedom and $\theta,\phi$ characterize the direction of the magnetization before rotation, i.e. $\mathbf{m}=(\sin \theta \cos \phi ,\, \sin \theta  \sin \phi ,\, \cos \theta)$. In turn, we obtain $Q= Q_\parallel+ Q_\perp$ with $Q_\parallel=[\dot \phi (1-\cos \theta) - \dot \chi] \frac{\sigma_z}{2}$ and $Q_\perp = \exp(i \chi \sigma_z) (\dot \phi \sin \theta \frac{\sigma_x}{2} - \dot \theta \frac{\sigma_y}{2} ) \exp(i \phi \sigma_z)$.
The term $Q_\parallel$ is diagonal in the spin-space. It is induced by the angular motion of the magnetization and appears in the action, eq. \eqref{eq: SU2 rotated action}, as an additional spin-dependent energy, which can also be 
interpreted in terms of the Berry-phase \cite{PhysRevLett.114.176806}. The term $Q_\perp$ is also related to the angular motion of the magnetization. However, it is purely off-diagonal in the spin-space. Therefore, it is related to transitions of individual electrons between the spin-up and spin-down states, i.e. the Landau-Zener transitions \cite{PhysRevLett.114.176806}.

\subsection{$U(1)$ gauge transformations}
We split $M$ and $V_d$ into constant parts and small deviations, i.e. $M= M_0 + \delta M$ and $V_d=V_{d0} + \delta V_d$. To deal with those deviations, we perform two $U(1)$-gauge transformations analog to \cite{AES_PRL, AES_PRB, PhysRevB.95.075425}. We use $e^{i \eta \frac{\sigma_z}{2}}$ for the length of the magnetization and $e^{-i \psi}$ for the voltage. Together, we have,
\begin{equation}
U = e^{i \eta \frac{\sigma_z}{2}} e^{-i \psi}\ , \label{eq: U}
\end{equation}
and would like to choose $\dot \eta= \delta M$ and $\dot \psi = \delta V_d$ on the Keldysh contour such as to completely eliminate $\delta M$ and $\delta V_d$. This choice would lead to boundary conditions $\eta_-(-T_K) - \eta_+(-T_K) = \oint_K dt\, \delta M = \int_{-T_K}^{T_K} dt\, \delta M_q = \delta M_q(\omega = 0)\equiv 2 T_K\, \delta M^q_{0}$ and analogously $\psi_-(-T_K) - \psi_+(-T_K)  = \delta V_d^q(\omega = 0) \equiv 2T_K\, \delta V^q_{d0}$. Although this is possible in principle, it is technically easier to choose the gauges to satisfy the boundary conditions $\eta_-(-T_K)-\eta_+(-T_K)= 4 \pi k$ and $\psi_-(-T_K)-\psi_+(-T_K) = 2 \pi l$ with $k, l \in \mathbb{Z}$.
It is possible to find a compromise of both and choose the gauges \cite{PhysRevB.95.075425},
\begin{eqnarray}
\dot \eta_\pm\! & = &\! \delta M_\pm \mp \frac{1}{2} \delta M^q_0\ ,\\
\dot \psi_\pm\! & = &\! \delta V_d^\pm \mp \frac{1}{2} \delta V^q_{d0}\ ,
\end{eqnarray}
which satisfies the boundary conditions with $k=0$ and $l=0$ and eliminates all of $\delta M, \delta V_d$ but their quantum zero-modes $\delta M^q_0, \delta V^q_{d0}$.
For the action, we obtain,
\begin{equation}
i \mathcal S = \mathrm{tr}\, \mathrm{ln} \left[ G_{z}^{-1} + \frac{\delta M^q_0}{2} \frac{\sigma_z}{2} - \frac{\delta V^q_{d0}}{2} - D^\dagger \Sigma D - \tilde Q \right] + i \mathcal{S}_\mathrm{HS}\, ,\label{eq: gauge transformed action}
\end{equation}
with $G_{z}^{-1}=G_0^{-1} +  \frac{M_0}{2} \sigma_z- V_{d0}$ and
the combined $U(1) \otimes U(1) \otimes SU(2)$-gauge-transformation,
\begin{equation}
D= R\, U\ ,
\end{equation}
where $R$ is the $SU(2)$-gauge transformation defined in eq. \eqref{eq: R} and $U$ stands for the combined $U(1)\times U(1)$-gauge transformation, eq. \eqref{eq: U}. Furthermore, $\tilde Q = Q_\parallel + \tilde Q_\perp$ is the transformed $Q$ with $\tilde Q_\perp = e^{- i \frac{\eta}{2} \sigma_z} Q_\perp  e^{i \frac{\eta}{2} \sigma_z}$, which is still purely off-diagonal in spin-space. $Q_\parallel$ is not affected by the $U(1)$ gauge transformations, since it is local in time-space and diagonal in spin-space.

Eq. \eqref{eq: gauge transformed action} is still formally exact\footnote{We did not yet make use of the approximate form of the self-energy}, but this is as far, as we can go without approximation. Now, we set out to derive the quasi-classical equations of motion for the magnetization and electrical potential jointly.

\subsection{Quasiclassical approximation \\ Expansion of the action in quantum components}
In principle, a straightforward variation with respect to the quantum fields directly leads to the (noiseless) quasiclassical equations of motion\footnote{The presence of a noise term can lead to a drift term changing the deterministic dynamics. One should keep in mind, that, in this sense, 'noiseless' is not equivalent to 'deterministic'. Further, we want to note that it is a different question, if the dynamics of a system is well described by quasi-classical equations of motion, e.g. instanton-like effects might influence the dynamics.}. In practice, however, this procedure leads to complicated integral or integro-differential equations, whose exact solution is usually out of reach. So, to gain insight into the dynamics, approximations have to be made. It is important, however, to first expand in quantum components and only afterwards in other small quantities. In particular, would we expand in tunneling before the expansion in quantum components, the important information about the electron distribution function on the dot could be lost \cite{PhysRevB.82.155317}.

For the purpose of expanding in quantum components, 
we perform the standard Keldysh rotation from the $(+,-)$ basis to the $(c,q)$ basis (note that for zero frequency 
components $\delta M_0^q$ and $\delta V_{d0}^q$ this has been already done in the previous subsection).
We introduce purely classical transformations $D_k = D|_{q=0}=D_c|_{q=0}$, where $...|_{q=0}$ means setting the quantum components of all coordinates to zero (note that $D_c\equiv (D_+ + D_-)/2$ is not equal to $D_k$ if the quantum components of the dynamical variables do not vanish\footnote{$D_{\pm}\equiv D(\phi_{\pm},\theta_{\pm},\chi_{\pm},\eta_{\pm},\psi_{\pm})$}). Then, we separate the purely classical part of the rotated self-energy\footnote{We note that $\left[D_c^\dagger \Sigma D_c\right]|_{q=0} = D^\dagger_k \Sigma D_k$.} from the rest $D^\dagger \Sigma D = D^\dagger_k \Sigma D_k + \delta \Sigma$. We proceed analogously for $\tilde Q= \tilde Q_k + \delta \tilde Q$, where $\tilde Q_k =\tilde Q_c|_{q=0}$. Then, all terms in $\delta \Sigma$ and $\delta \tilde Q$ are at least of first order in quantum components.
For the action we obtain,
\begin{equation}
i \mathcal S = \mathrm{tr}\, \mathrm{ln} \left[ G^{-1} + \frac{\delta M^q_0}{2} \frac{\sigma_z}{2} - \frac{\delta V^q_{d0}}{2} - \delta \Sigma - \frac{\delta \tilde Q}{2} \right] + i \mathcal{S}_\mathrm{HS}\, ,
\end{equation}
where we have absorbed $\tilde Q_k$ and $D_k \Sigma D_k$ into the classical Green's function $G_c$ defined by, 
\begin{equation}
G_c^{-1}= G_z^{-1} - \tilde Q_k - D_k^\dagger \Sigma D_k\ . \label{eq: inverse GF}
\end{equation}
We emphasize that $G_c$ is not the full Green's function of the dot. Instead, it is of an auxiliary character, since only the purely classical parts of the rotation-, length-, and potential-dynamics are included. Furthermore, it is a Green's function in the rotating frame.

We can now expand the action in quantum components, i.e. in $\delta M_0^q$, $\delta V_{d0}^q$, $\delta \Sigma$, and $\delta \tilde Q$. 

To first order in $\delta M_0^q$ and $\delta V_{d0}^q$, we obtain the zero-mode (zm) contributions to the action,
\begin{eqnarray}
i \mathcal{S}_{zm}^M & = & \frac{1}{4} \mathrm{tr} \left[ G_c\, \delta M_0^q\, \sigma_z \right]\ , \label{eq: ZMM} \\
i \mathcal{S}_{zm}^V & = & - \frac{1}{2} \mathrm{tr} \left[ G_c\, \delta V_{d0}^q\right]\ , \label{eq: ZMV}
\end{eqnarray}
which will turn out to be important for the determination of $M_0$ and $V_{d0}$.

Analog to $\tilde Q$, we split the contribution of $\delta \tilde Q$ into two, i.e. $\delta \tilde Q =\delta Q_\parallel + \delta \tilde Q_\perp$, where $\delta Q_\parallel$ is purely spin-diagonal and $\delta \tilde Q_\perp$ is purely spin-off-diagonal. To first order in $\delta Q_\parallel$, we obtain an action of the Wess-Zumino-Novikov-Witten type,
\begin{equation}
i \mathcal{S}_\mathrm{WZNW} = - \frac{1}{2} \mathrm{tr} \left[G_c\, \delta Q_\parallel \right]\ , \label{eq: original WZNW}
\end{equation}
which describes the contribution of the Berry-phase. To first order in $\delta \tilde Q_\perp$, we obtain,
\begin{equation}
i \mathcal{S}_\mathrm{LZ} = - \frac{1}{2} \mathrm{tr} \left[G_c\, \delta \tilde Q_\perp\right]\ , \label{eq: LZ action}
\end{equation}
which is related to Landau-Zener transitions \cite{PhysRevLett.114.176806}. 

To first order in $\delta \Sigma$, we obtain an Ambegaokar-Eckern-Sch\"on-like action \cite{AES_PRL,AES_PRB},
\begin{equation}
i \mathcal S_\mathrm{AES} = - \mathrm{tr} \left[G_c\, \delta \Sigma \right]\ , \label{eq: original AES}
\end{equation}
which carries information about effects related to tunneling. In particular, it contains information about currents and dissipation.

Before we can obtain an explicit form of the effective action, we have to determine the classical Green's function $G_c$.

\section{Determination of the classical Green's function \label{sec: determine GF}}
The classical Green's function $G_c$ has to be determined from its inverse, defined in eq. \eqref{eq: inverse GF}. This corresponds to solving a kinetic equation. While it is rather straightforward to invert $G_z^{-1}$, the dependence of $\tilde Q_k$ and $D_k^\dagger \Sigma D_k$ on the trajectories of $\mathbf M$ and $V_d$ can create quite complicated time-dependence. Thus, for arbitrary trajectories of $\mathbf M$ and $V_d$ this poses a very hard problem. We do not attempt to solve this problem in its full generality. Instead, we present a strategy for the dot being deep in the Stoner-regime, with a large magnetization $M_0$. At first, following the ideas of ref. \cite{PhysRevLett.114.176806}, we perform an adiabatic approximation and use a specific choice of gauge $\chi$ to deal with the term $\tilde Q_k$. Afterwards, we employ the slowness of coordinates $\theta_c, \dot \phi_c$ to deal with the rotated self-energy $D_k^\dagger \Sigma D_k$.

\subsection{Ferromagnetic regime, adiabatic approximation and choice of gauge} 
We assume the dot to be deep in the Stoner-regime. Then, thinking in terms of Landau-theory of phase transitions\footnote{Note that in the non-equilibrium situation the Landau-theory serves only as a guiding idea.}, there is a well established minimum for the length of the magnetization $M_c$. This means that the dynamic length fluctuations $\delta M_c$ around the large, but constant, value $M_0$ are small $\delta M_c \ll M_0$.

The magnetization length $M_0$ is assumed to be the largest relevant energy scale in the dot. The classical Green's function $G_c$ has to be determined from its inverse, eq. \eqref{eq: inverse GF}, where $M_0$ appears only in the spin-diagonal components with different signs for the spin-up and spin-down components. Therefore, the diagonal elements of $G_c^{-1}$ are never degenerate and, thus, the spin-off-diagonal elements of $G_c$ are suppressed by $1/M_0$. To leading order in $\frac{1}{S}$, we can disregard the spin-off-diagonal parts of both $\tilde Q_k$ and $D_k^\dagger \Sigma D_k$ when calculating the classical Green's function $G_c$. This means we disregard $\tilde Q_\perp^k$, i.e., the Landau-Zener-transitions, which corresponds to the adiabatic approximation. Expressed in more physical terms, the dynamics of the direction of magnetization $\mathbf m$ is very slow compared to the time scale related to the length of magnetization $M$, such that spins of individual electrons adiabatically follow $\mathbf m$. Thus, Landau-Zener transitions can be disregarded  \cite{PhysRevLett.114.176806}.

The part $Q_\parallel^k$ remains, even in the adiabatic approximation, since it is diagonal in spin-space. However, to deal with this contribution, we employ the gauge freedom $\chi$ as is done in ref. \cite{PhysRevLett.114.176806}. That is, we eliminate of $Q_\parallel^k$ while simultaneously respecting the boundary conditions on the Keldysh contour $\chi_-(-T_K) - \chi_+(-T_K) = 4\pi n$ with $n \in \mathbb Z$. This is achieved by \cite{PhysRevLett.114.176806},
\begin{eqnarray}
\dot \chi_c & = & \dot \phi_c (1-\cos \theta_c)\ , \\
\chi_q & = & \phi_q (1- \cos \theta_c)\ .
\end{eqnarray}
Then, up to first order in quantum components, we obtain $\delta Q_\parallel= \sin \theta_c ( \dot \phi_c \theta_q - \dot \theta_c \phi_q) \frac{\sigma_z}{2}$.

To summarize: $Q_\parallel^k$ is eliminated by a choice of gauge $\chi$ and $\tilde Q_\perp^k$ can be disregarded in adiabatic approximation. This reduces equation \eqref{eq: inverse GF} for the inverse classical Green's function to,
\begin{equation}
G_c^{-1}=G_z^{-1} - D_k^\dagger \Sigma D_k\ . \label{eq: inverse GF without Q}
\end{equation}
The rotated self-energy $D_k^\dagger \Sigma D_k$ will be treated next. We keep in mind that, due to $M_0$ being the largest relevant energy scale in the dot, the spin-off-diagonal parts will be negligible.
 
\subsection{Separation of time-scales}
Now, we make use of the fact that the dynamics take place at various time-scales.

We define a coordinate to be slow, if it changes on time-scales $\tau_{coord.} \gg \mathrm{max}( \tau_\Gamma , \tau_{T})$, where the life-time of electrons in the dot $\tau_\Gamma = \frac{1}{\Gamma}$ with a generic tunneling rate $\Gamma$; and the correlation time of thermal noise $\tau_{T}\equiv \frac{1}{T}$ with $T\equiv  \mathrm{min}(T_l , T_r)$. According to this definition, the distribution function adjusts adiabatically to changes in slow coordinates, since the life-time of electrons determines the time-scale at which the distribution function can react to changes. Furthermore, the thermal noise appears to be white for slow coordinates. These facts allow for a simplified treatment of slow coordinates, by making use of a gradient expansion.
For that purpose, we define a slow gauge transformation $D_s$ which originates from $D_k$ by keeping all slow coordinates for which we want to exploit the slowness and simply setting all other coordinates to zero. 
Then, in eq. \eqref{eq: inverse GF without Q}, we subtract and add the slowly rotated self-energy $D_s^\dagger \Sigma D_s$,
\begin{equation}
G_c^{-1}=G_z^{-1} - D_s^\dagger \Sigma D_s - (D_k^\dagger \Sigma D_k - D_s^\dagger \Sigma D_s)\ ,
\end{equation}
and expand in the difference between purely classical rotated self-energy and the slowly rotated self-energy $(D_k^\dagger \Sigma D_k - D_s^\dagger \Sigma D_s)$. It follows,
\begin{equation}
G_c = G_s + G_s (D_k^\dagger \Sigma D_k - D_s^\dagger \Sigma D_s) G_s + ...\ , \label{eq: expansion of GF}
\end{equation}
with the slow Green's function $G_s$ defined by,
\begin{equation}
G_s^{-1}=G_z^{-1} - D_s^\dagger \Sigma D_s\ , \label{eq: inverse slow GF}
\end{equation}
The gain of this procedure is that the slow Green's function $G_s$ can be determined approximately by use of a gradient expansion, App. \ref{app: gradient exp new}. Contributions to the classical Green's function from the other coordinates (not included in $D_s$) are found by expansion, eq. \eqref{eq: expansion of GF}. We emphasize that it is optional for a slow coordinate to either include it into $D_s$ and exploit its slowness, or to proceed on more general grounds with the expansion, eq. \eqref{eq: expansion of GF}. 

Next, to be more explicit, we consider the time-scales of the actual coordinates of the model system.

Deep in the Stoner-regime, with a large magnetization $M_0$, the coordinates $\theta_c$ and $\dot \phi_c$ are slow. The reason is that both, $\theta_c$ and $\dot \phi_c$ change only due to tunneling of electrons.
According to simple geometrical arguments, those changes are suppressed by the length of the magnetization $M_0$, respectively the spin $S$. Thus, we expect $\tau_\theta, \tau_{\dot \phi} \propto \frac{S}{\Gamma}$ and in turn $\tau_\theta, \tau_{\dot \phi} \gg \mathrm{max}(\tau_\Gamma, \tau_{T})$, if temperatures are not too low. 
We emphasize a subtle but important point: It is $\dot \phi$ which must be slow; not $\phi$ itself. The magnetization will precess around the external magnetic field roughly with the frequency determined by the external magnetic field $B$. The effects of this precession are particularly interesting, if the precession frequency is larger than the level broadening $B \gg \Gamma_\sigma(\theta)$. Then, however, $\phi$ is not a slow variable, whereas $\dot \phi$ still is.

Also the electrical potential $\delta V_d^c$ and length of the magnetization $\delta M_c$ change only due to tunneling. However, there is no geometric suppression for those. We expect $\tau_{\delta V_d^c} \propto \frac{1}{\Gamma}$ and $\tau_{\delta M_c} \propto \frac{1}{\Gamma}$. Therefore, we cannot assume $\delta M_c$ and $\delta V_d^c$ to be slow variables. Indeed, $\delta V_d^c$ turns out to be fast compared to changes in the distribution function, i.e. $\tau_{\delta V_d} \ll \tau_\Gamma$, while $\delta M_c$ will typically\footnote{Depending on the details of the density of states, it might happen that $\delta M_c$ is also a slow variable. However, even we would not include it into $D_s$.}
 change on a time-scale similar to that of the distribution function $\tau_{\delta M} \approx \tau_\Gamma$, details are provided in App. \ref{app: full dynamics}. 

Furthermore, due to the large spin $S$, we observe a separation of time-scales $\tau_{\theta}, \tau_{\dot \phi} \gg \tau_{\delta V_d}, \tau_{\delta M}$ for the coordinates. Both $\delta M_c$ and $\delta V_d^c$ will almost immediately relax to zero on the typical time-scale of the angular dynamics.
Being mainly interested in the angular dynamics, this justifies to disregard $\delta M_c$ and $\delta V_d^c$ (resp. $\eta, \psi$), as we will do in the main text. However, due to its interplay with the dynamic distribution function, the treatment of $\delta M_c$ poses an interesting technical problem by itself. This is solved in App. \ref{app: full dynamics} as part of the full problem with all four coordinates.

\subsection{The slow Green's function}
We employ the slowness of angular coordinates $\theta_c, \dot \phi_c$, now, by setting $D_s=R_k$, where $R_k = R_c|_{q=0}$. Then, for the slow Green's function it follows,
\begin{equation}
G^{-1}_s = G_{z}^{-1}  -  R^\dagger_k  \Sigma  R_k\ . \label{eq: definition of inverse slow GF}
\end{equation}
Using the slowness of $R_k$, we can determine the rotated self-energy $R^\dagger_k  \Sigma  R_k$ approximately, see App. \ref{app: slowly rotated self energy}. Then, we perform a gradient expansion, see App. \ref{app: gradient exp new}, and keep the zeroth-order only. Using the Wigner time/frequency coordinates $(\cmt,\omega)$ (see App. \ref{app: gradient exp new}) we obtain
\begin{eqnarray}
&&G_s^{R/A}(\cmt,\omega)= \frac{1}{\omega - \xi_{\alpha \sigma}  \pm i \Gamma_\sigma (\theta_c)}\ , \label{eq: main slow retarded GF} \\
&&\hspace{-1em} G_s^{K}(\cmt,\omega)\! =\!\frac{  -2i\, \Gamma_\sigma (\theta_c)}{(\omega - \xi_{\alpha \sigma} )^2\! +\! \Gamma_\sigma^2 (\theta_c)} F_s^\sigma (\cmt,\omega)\ , \label{eq: main slow keldysh GF}
\end{eqnarray}
with $\xi_{\alpha \sigma} = \epsilon_\alpha + V_{d0} - \frac{M_0}{2} \sigma$ which denote the single-particle energy for level $\alpha$ and spin $\sigma$, where the (stationary) mean-fields $V_{d0}, M_0$ are included. Further, we introduced the level broadening $\Gamma_\sigma (\theta_c) = \cos^2 \frac{\theta_c}{2} \Gamma_l^\sigma + \sin^2 \frac{\theta_c}{2} \Gamma_l^{\bar{\sigma}} + \Gamma_r$, where $\bar{\sigma}$ is the spin value opposite to $\sigma$ and $\theta_c = \theta_c(\cmt)$. The slow distribution function is given by, 
\begin{eqnarray}
\hspace{-2em}F_s^\sigma (\cmt,\omega)= \frac{1}{\Gamma_\sigma (\theta_c)}\!\! &\bigg[&\!\!\cos^2 \frac{\theta_c}{2}\, \Gamma_l^\sigma\, F_l\left(\omega + \sigma\, \omega_- \right) + \nonumber \\ 
&+&\! \sin^2 \frac{\theta_c}{2}\, \Gamma_l^{\bar \sigma}\, F_l\left(\omega + \bar \sigma\,  \omega_+ \right) + \nonumber \\
&+&\! \cos^2 \frac{\theta_c}{2}\, \Gamma_r\, F_r\left(\omega + \sigma\,  \omega_- \right) + \nonumber \\
&+&\! \sin^2 \frac{\theta_c}{2}\, \Gamma_r\, F_r\left(\omega + \bar \sigma\, \omega_+ \right) \bigg] \ , \label{eq: non-eq. distr.}
\end{eqnarray}
where $F_{l/r}(\omega)= \tanh \frac{\omega - \mu}{2\, T_{l/r}}$ and the Berry-phase enters through the dynamic shifts $\omega_\pm  = \dot \phi_c(\cmt) (1 \pm \cos \theta_c(\cmt))/2$. The distribution function $F_s^\sigma (\cmt,\omega)$ is a superposition of four different equilibrium distribution functions and therefore is clearly a non-equilibrium distribution. In Fig. \ref{fig: stationary} (b) the distribution function $n_s^\sigma(\cmt,\omega) = [1-F_s^\sigma(\cmt,\omega)]/2$ is shown for spin-up electrons for two persistent precessions at different stationary angles $\theta_c(\cmt)=\theta_0$.

\section{Quasiclassical equations of motion \label{sec: EOM}}
We use the slow Green's function and determine the contributions to the effective action. Afterwards, we vary the action with respect to the quantum components $\theta_q, \phi_q$ to obtain the quasi-classical equations of motion.

\subsection{Effective action for slow dynamics}
The determination of the Hubbard-Stratonovich decoupling contribution, eq. \eqref{eq: original HS}, is straightforward and we obtain,
\begin{equation}
i \mathcal S_\mathrm{HS} = - i \frac{M_0 B}{2J} \int dt\, \theta_q \sin \theta_c\ , \label{eq: HS linearized angles only}
\end{equation}
where we used $\delta M= 0$, $\delta V_d = 0$ and dropped constant terms. 

The zero-mode contributions, eqs. \eqref{eq: ZMM} \eqref{eq: ZMV}, are not directly relevant for the angular dynamics, only for $M_0$ and $V_{d0}$, see App. \ref{sec: zero modes}.

For the slow part of the WZNW action, eq. \eqref{eq: original WZNW}, we obtain,
\begin{equation}
i \mathcal S_\mathrm{WZNW}
= - i \int dt\, S \sin \theta_c\, (\theta_q \dot \phi_c - \phi_q \dot \theta_c)\ , \label{eq: linearized WZNW angles}
\end{equation}
where we have explicitly taken the trace over time- and Keldysh-space and introduced,
\begin{eqnarray}
S\!\! & = &\!\! -\frac{i}{2} \mathrm{tr}\left[G_s^<(t,t) \sigma_z\right] \label{eq: slow spin} \\
& = &\!\! -\frac{1}{4} \int d\omega\, \left[ \rho_\uparrow (\omega) F_s^\uparrow (t, \omega) - \rho_\downarrow (\omega) F_s^\downarrow (t, \omega) \right]\ , \label{eq: WZNW-spin}
\end{eqnarray}
with the density of states $\rho_\sigma(\omega)= \sum_\alpha  \frac{1}{\pi} \frac{\Gamma_\sigma (\theta_c)}{(\omega - \xi_{\alpha \sigma})^2 + (\Gamma_\sigma (\theta_c))^2}$, which is broadened by $\Gamma_\sigma(\theta_c)$ and shifted by $\sigma M_0/2-V_{d0}$. We note that $S$ is the length of the spin, i.e. it is half the difference of the number of spin-up and spin-down electrons on the dot.

The LZ-action, eq. \eqref{eq: LZ action}, vanishes in the approximation for a spin-diagonal slow Green's function, since $\delta \tilde Q_\perp$ is purely spin-off-diagonal.

We split the AES-like action, eq. \eqref{eq: original AES}, into a retarded part containing all terms of first order in $R_q$ and the rest. The rest, which includes the Keldysh part (second order in $R_q$), is at least of second order in quantum components. Therefore, it only contributes to noise which will be studied in future work. For the noiseless dynamics, studied here, it is sufficient to know the retarded part,
\begin{equation}
i \mathcal{S}_\mathrm{AES}^{R}\! =\! - i\!\! \int\!\! dt\, dt' \sum_{\sigma \sigma'} \mathrm{Im}\! \left[ R_q^{\sigma' \sigma}(t)\, \alpha_{s,\sigma \sigma'}^R(t,t')\, (R_{c}^{\sigma' \sigma}(t'))^* \right]\, , \label{eq: ret. AESlike action}
\end{equation}
where we have explicitly taken the trace over time-, Keldysh-, and spin-space and used $(R_{c}^{\sigma' \sigma}(t'))^*= (R^\dagger_c(t'))_{\sigma \sigma'}$. The slow retarded kernel function is defined by,
\begin{equation}
\alpha_{s,\sigma \sigma'}^R(t,t')\! =\! \mathrm{tr}\! \left[G_{s \sigma}^R(t,t') \Sigma^K_{\sigma'}(t'\!\!-\!t)\! +\! G_{s \sigma}^K(t,t') \Sigma^A_{\sigma'}(t'\!\!-\!t) \right]\, .\label{eq: retarded kernel}
\end{equation}
We note that in order to obtain eqs. \eqref{eq: ret. AESlike action}, \eqref{eq: retarded kernel} we have split $\delta \Sigma$ apart. The dynamical fields, contained in $R_q$ and $R_c$, are written separately from the unrotated self-energy $\Sigma_\sigma(t'-t)$, which is included in the kernel function, eq. \eqref{eq: retarded kernel}.

We can now proceed by calculating the retarded kernel function:
\begin{eqnarray}
\alpha^R_{s, \sigma \sigma'}(\cmt,\omega)\! =\!\! \int\! d\omega'\, & \rho_\sigma(\omega')\,& \Big[
\Gamma_l^{\sigma'}\! \left(F_{s}^\sigma(\cmt,\omega')- F_l(\omega'-\omega)\right)  \nonumber \\ &  & \hspace{-0.5em}+ \Gamma_r \left(F_{s}^\sigma(\cmt,\omega')- F_r(\omega'-\omega)\right)\Big]\ , \nonumber \\
\end{eqnarray}
where we disregarded the imaginary part, since we expect it to only renormalize the external magnetic field.
We further assume the shifted density of states to be approximately linear around the electrochemical potential $\mu$, i.e. $\rho_\sigma(\mu + \omega) \approx \rho_\sigma + \rho_{\sigma}' \omega$, with $\rho_\sigma=\rho_\sigma(\omega = \mu )$ and $\rho_{\sigma}'= [\partial_\omega \rho_\sigma(\omega)]_{\omega = \mu }$, on all relevant scales less than $M_0$. In particular it should be approximately linear on the scale of temperatures $T_{l/r}$. We assume that the density of states changes roughly on the scale of the magnetization, thus, the derivative of the density of states is roughly of the order $\mathcal{O}(1/S)$. We will only keep those terms with $\rho_{\sigma}'$ that also include the temperatures, which can be made large enough to compensate the smallness of $\rho_{\sigma}'$. We obtain,
\begin{equation}
\alpha^R_{s,\sigma \sigma'}(\cmt,\omega)=I^{\sigma \sigma'}_\mathrm{h}(\theta_c, \dot \phi_c) + I^{\sigma \sigma'}_d(\theta_c) + g_{\sigma \sigma'} \omega\ , \label{eq: alpha ret. explicit}
\end{equation}
where $\theta_c= \theta_c(\cmt)$, $\dot \phi_c= \dot \phi_c(\cmt)$ and we introduced the conductances $g_{\sigma \sigma'} = 2 \rho_\sigma (\Gamma_l^{\sigma'} + \Gamma_r)$ in the dissipative contribution and the current related to thermal driving (thermoelectric effect) $I^{\sigma \sigma'}_d(\theta_c)=\frac{\Gamma_r \Gamma_\Delta}{\Gamma_\sigma(\theta_c)}  (\sigma' - \sigma \cos \theta_c) \rho_{\sigma}' d$, where $d= \frac{\pi^2}{3} (T_l^2 - T_r^2)$ is a parameter describing the thermal driving and $\Gamma_\Delta = (\Gamma_l^\uparrow - \Gamma_l^\downarrow)/2$. 
Further, we introduced a "hybrid"-current related to the precession of the magnetization (geometric phase)  $I^{\sigma \sigma'}_\mathrm{h}(\theta_c, \dot \phi_c)= g_{\sigma \sigma'}  \frac{\Gamma_\Delta \sin^2 \theta_c}{2 \Gamma_\sigma(\theta_c)}  \dot \phi_c$. 
This current arises due to the effect of precession on the distribution function of the dot.
Its name will become clear, when we discuss the equations of motion.  

It is now straightforward to insert the retarded kernel function, eq. \eqref{eq: alpha ret. explicit}, into the retarded AES-like action, eq. \eqref{eq: ret. AESlike action}. To first order in quantum components, we obtain the explicit result,
\begin{equation}
i \mathcal{S}_\mathrm{AES}^{R}\! =\! - i\!\! \int\!\! dt\, \left\lbrace \theta_q \tilde g(\theta) \dot \theta + \phi_q \sin^2 \theta\left[ \tilde g(\theta) \dot \phi - I_s(\theta, \dot \phi) \right]\right\rbrace \ , \label{eq: linearized AES angles}
\end{equation}  
where $\theta = \theta_c(t), \dot \phi= \dot \phi_c(t)$ and the function $\tilde g(\theta) = \frac{g_{\uparrow \uparrow} + g_{\downarrow \downarrow}}{4} \sin^2\frac{\theta}{2} + \frac{g_{\uparrow \downarrow} + g_{\downarrow \uparrow}}{4} \cos^2\frac{\theta}{2}$ has dimensions of conductance and is responsible for angular dissipation~\cite{ChudnovskiyPRL,KamenevBook}. Further, we defined the spin-transfer-torque (STT) current $I_s(\theta, \dot \phi)= I_h^s(\theta, \dot \phi) + I_d^s(\theta)$ with two contributions: a thermal one $I_{d}^s(\theta)=\frac{1}{4}[I_{d}^{\uparrow \uparrow}(\theta) - I_{d}^{\uparrow \downarrow}(\theta) + I_{d}^{\downarrow \uparrow}(\theta) - I_{d}^{\downarrow \downarrow}(\theta)]$ and a hybrid-STT-current $I_{h}^s(\theta, \dot \phi)=\frac{1}{4}[I_{h}^{\uparrow \uparrow}(\theta, \dot \phi) - I_{h}^{\uparrow \downarrow}(\theta, \dot \phi) + I_{h}^{\downarrow \uparrow}(\theta, \dot \phi) - I_{h}^{\downarrow \downarrow}(\theta, \dot \phi)]$ related to the precession of the magnetization.

\subsection{Landau-Lifshitz-Gilbert-Slonczewski equation}
The variation of the action consisting of $i \mathcal S_\mathrm{HS}$, $i \mathcal S_\mathrm{WZNW}$, and $i \mathcal S_\mathrm{AES}$ from eqs. \eqref{eq: HS linearized angles only}, \eqref{eq: linearized WZNW angles}, and \eqref{eq: linearized AES angles} with respect to quantum components is straightforward and yields the quasi-classical equations of motion,
\begin{eqnarray}
\sin \theta\, \dot \phi & = & - \sin \theta\, B - \frac{\tilde g(\theta)}{S} \dot \theta\ , \label{eq: 0eom for phi}\\
\sin \theta\, \dot \theta & = & \frac{\sin^2 \theta}{S} \left[ \tilde g(\theta) \dot \phi - I_h^s(\theta, \dot \phi) - I_d^s(\theta)\right]\ . \label{eq: 0eom for theta}
\end{eqnarray}
For simpler notation, we suppress the index for classical components here and in the following\footnote{After the variation, there are no more quantum components, i.e. all coordinates are classical and no confusion can arise.}. For the spin-transfer torque currents, we obtain explicitly,
\begin{eqnarray}
I_h^s(\theta, \dot \phi) & = & \frac{\Gamma_\Delta^2 \sin^2 \theta}{\Gamma_\uparrow(\theta) \Gamma_\downarrow(\theta)}  \tilde g(\theta)  \dot \phi \ ,\\
I_d^s(\theta) & = & \frac{\Gamma_r \Gamma_\Delta}{\Gamma_\uparrow(\theta) \Gamma_\downarrow(\theta)}  \tilde g'(\theta) d\ ,
\end{eqnarray}
and we defined $\tilde g'(\theta)=\frac{g'_{\uparrow \uparrow} + g'_{\downarrow \downarrow}}{4} \sin^2\frac{\theta}{2} + \frac{g'_{\uparrow \downarrow} + g'_{\downarrow \uparrow}}{4} \cos^2\frac{\theta}{2}$ with $g_{\sigma \sigma'}' = 2 \rho_\sigma' (\Gamma_l^{\sigma'} + \Gamma_r)$. For convenience, we restate the previous definitions $\Gamma_\sigma(\theta)= \Gamma_l^\uparrow \cos^2 \frac{\theta}{2}  + \Gamma_l^\downarrow \sin^2 \frac{\theta}{2} + \Gamma_r$ and $\tilde g(\theta)=\frac{g_{\uparrow \uparrow} + g_{\downarrow \downarrow}}{4} \sin^2\frac{\theta}{2} + \frac{g_{\uparrow \downarrow} + g_{\downarrow \uparrow}}{4} \cos^2\frac{\theta}{2}$ and $\Gamma_\Delta = (\Gamma_l^\uparrow - \Gamma_l^\downarrow)/2$. Defining further $\rho_{\Sigma / \Delta} \equiv \rho_\uparrow \pm \rho_\downarrow$, $\rho_{\Sigma / \Delta}' \equiv \rho_{\uparrow}' \pm \rho_{\downarrow}'$, and $\Gamma_\Sigma \equiv\frac{1}{2} (\Gamma_l^\uparrow +\Gamma_l^\downarrow) + \Gamma_r$ we can rewrite
$\Gamma_\sigma(\theta) = \Gamma_\Sigma+\sigma \Gamma_\Delta \cos\theta$, 
$\tilde{g}(\theta) = \frac{1}{2} \bigl ( \rho_\Sigma \Gamma_\Sigma - \rho_\Delta \Gamma_\Delta\cos\theta)$, 
and $\tilde{g}^\prime(\theta) = \frac{1}{2} \bigl ( \rho^\prime_\Sigma \Gamma_\Sigma - \rho^\prime_\Delta \Gamma_\Delta\cos\theta)$.

It is possible to recast the equations of motion \eqref{eq: 0eom for phi} and \eqref{eq: 0eom for theta} into a single equation of motion for the direction of the magnetization $\mathbf m$. We obtain the Landau-Lifshitz-Gilbert-Slonczewski (LLGS) equation \cite{slonczewski1996current}, 
\begin{equation}
\mathbf{\dot{m}} = \mathbf{m} \times \mathbf{B}  - \alpha(\theta)\, \mathbf{m}  \times \mathbf{\dot{m}} + \frac{1}{S} \mathbf{m} \times ( \mathbf{I}_s(\theta,\dot \phi) \times \mathbf{m} )\ ,
\end{equation}
where we used $\frac{M_0}{2J} \approx S$, see App. \ref{sec: zero modes}, and defined the Gilbert damping coefficient $\alpha(\theta)= \frac{\tilde g(\theta )}{S}$ and the direction of the STT-current is determined by the fixed magnetization $\mathbf{I}_s(\theta, \dot \phi) \parallel \mathbf{M}_\mathrm{fix}$, its magnitude is given by $I_s(\theta, \dot \phi)= I_h^s(\theta, \dot \phi) + I_d^s(\theta)$.

\subsection{Persistent precessions and the hybrid current}
We investigate the persistent precessions, i.e. solutions to the LLGS-equation, which precess around the external magnetic field at some frequency $\dot \phi=\omega_\mathrm{prec}$ at a constant angle $\theta=\theta_0$. 
For the system to support persistent precessions at a (non-trivial) angle $\theta_0\neq 0,\pi$, there has to be a balance of Gilbert-damping and STT-excitation. That is in eq. \eqref{eq: 0eom for theta} there must be a balance between dissipation $\tilde g(\theta) \dot \phi$, thermal STT-driving $- I_d^s(\theta)$ and the hybrid current $-I_h^s(\theta, \dot \phi)$. Note that the hybrid current is proportional to the precession frequency $\dot \phi$. This is the origin of its interesting hybrid role: While it is a contribution to the STT-current, it acts like a renormalization of the damping.

To determine the persistent precessions and their stability, we use the ansatz $\phi = \omega_\mathrm{prec} t + \delta \phi$ and $\theta = \theta_0 + \delta \theta$, with $\omega_\mathrm{prec}$ and $\theta_0$ constant. The persistent precessions are then found for $\delta \phi, \delta \theta = 0$. Their stability is determined by the dynamics of $\delta \theta$ only, since $\delta \phi$ turns out to be a marginal coordinate. If $\delta \theta$ relaxes towards zero, then we call the corresponding persistent precession stable; if $\delta \theta$ tends to grow away from zero, we call the corresponding persistent precession unstable.

From eq. \eqref{eq: 0eom for phi}, we immediately obtain the percession frequency $\omega_\mathrm{prec}=-B + \mathcal{O}(1/S^2)\approx -B$. Using this in eq. \eqref{eq: 0eom for theta}, we can determine the stationary polar angle $\theta_0$. There are always solutions at the poles $\sin \theta_0 = 0$, and other possible values are given by, 
\begin{equation}
\cos \theta_0 = \frac{\Gamma_\Sigma}{\Gamma_\Delta} \frac{\rho_\Sigma B + \lambda \rho_\Sigma' d}{\rho_\Delta B + \lambda \rho_\Delta' d}\ ,
\end{equation}
where $\lambda \equiv \Gamma_r \Gamma_\Delta / (\Gamma_\Sigma^2 - \Gamma_\Delta^2)$. This formula is, of course, only applicable, if the right hand side takes values between -1 and 1.

\begin{figure}
\begin{center}
\includegraphics[width=0.45\textwidth]{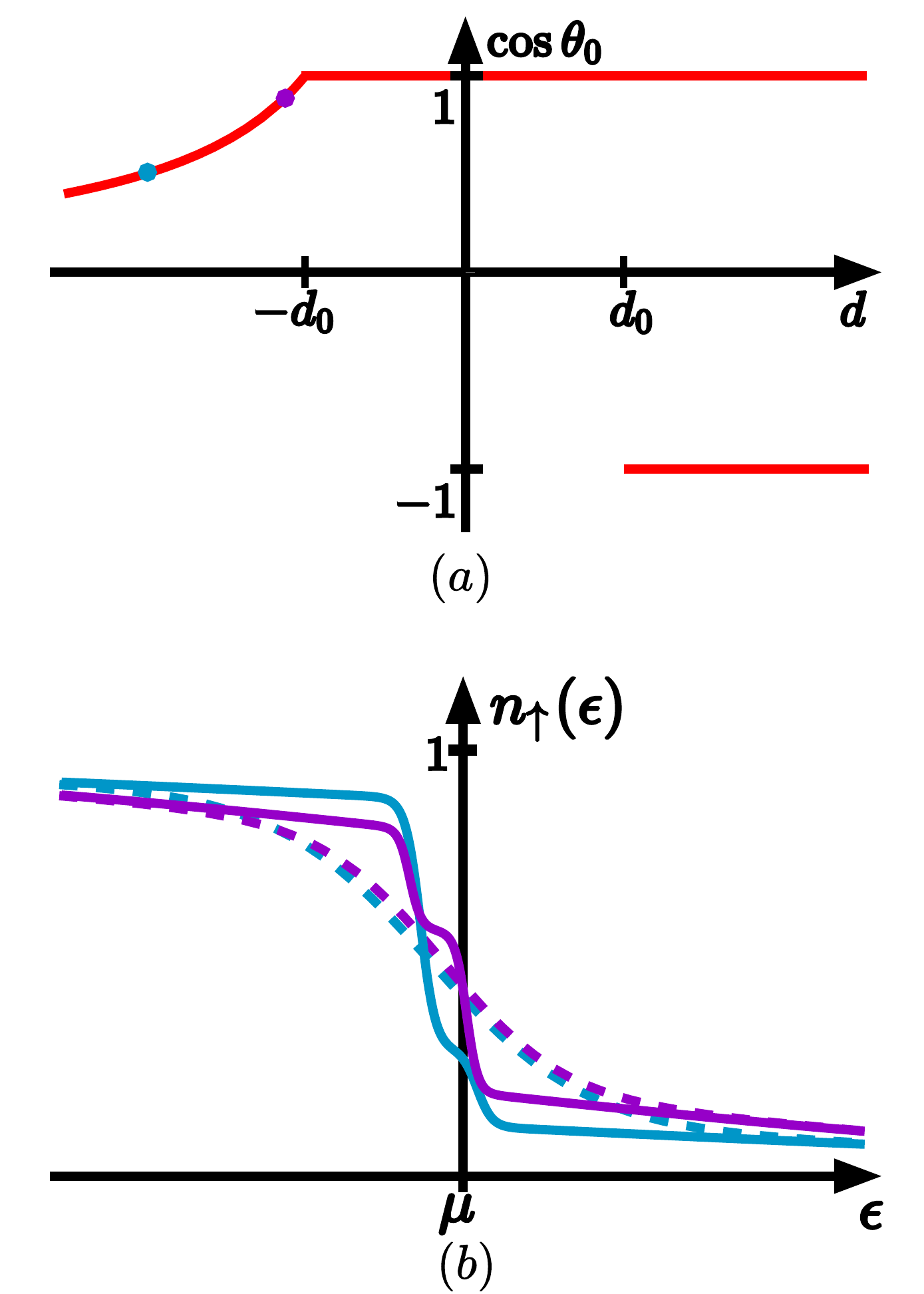}
\end{center}
\caption{For $d= \frac{\pi^2}{3} (T_l^2 - T_r^2)$ and $\Gamma_\Delta<0$, $\rho_\Delta'<0$ and a symmetric density of states, i.e. $\rho_\Delta =0$, $\rho_\Sigma'=0$, we show (a) the stationary solutions for $\cos \theta_0$ with their stability (red solid = stable, blue dotted = unstable) and (b) non-equilibrium distribution functions. The temperature difference tries to drive the magnetization towards the poles for $d>0$ and towards the equator for $d<0$. The Gilbert damping is stronger than thermal driving for $|d|<d_0$, where $d_0=-\Gamma_\Sigma \rho_\Sigma B / ( \lambda \Gamma_\Delta \rho_\Delta')$.}\label{fig: stationary}
\end{figure}

For a symmetric unshifted density of states, it follows $\rho_\Delta = 0$ and $\rho_\Sigma'=0$. For this density of states and with $\Gamma_\Delta<0$, $\rho_\Delta'<0$, we show stationary solutions for $\theta_0$ in Fig. \ref{fig: stationary} (a). The thermal driving ($d= \frac{\pi^2}{3} (T_l^2 - T_r^2)$) tries to drive the magnetization towards the poles for $d>0$ and towards the equator for $d<0$. However, the Gilbert damping is stronger than thermal driving for $|d|<d_0$, where $d_0=-\Gamma_\Sigma \rho_\Sigma B / ( \lambda \Gamma_\Delta \rho_\Delta')$. 
From Fig. \ref{fig: stationary} (a), we identify three regimes: For $-d_0 < d< d_0$ driving is too weak to compete with Gilbert damping and therefore the magnetization stays at the north-pole $\cos \theta_0=1$; For $d>d_0$ the south-pole becomes locally stable while at the northern hemisphere Gilbert damping and thermal driving cooperate and make the north-pole globally stable; For $d<-d_0$ the persistent precessions become stable for non-trivial angle $\theta_0$, which are determined by the mutual compensation of thermal driving and (renormalized) Gilbert damping.
In Fig. \ref{fig: stationary} (b) we show the distribution function on the magnet for the up-spins $n_\uparrow(\cmt, \omega) = (1- F_s^\uparrow(\cmt, \omega))/2$ in the rotating frame, for two persistent precessions ($\theta(\cmt) \rightarrow \theta_0$ and $\dot \phi (\cmt) \rightarrow - B $) marked in Fig. \ref{fig: stationary} (a). We emphasize that for a given driving parameter $d$, the distribution function is not unique. The solid and dashed lines are for the same driving parameter $d$ but different lead temperatures. While the non-equilibrium features of different lead temperatures $T_l, T_r$ and Berry-phase shifts $\omega_\pm$ can be clearly seen for the solid distributions, they are hidden, but not less relevant, for higher temperature $T_l$ for the dashed distributions.

\section{Enhancement of the thermoelectric effect by the pumping current \label{sec: currents}}
Finally, we consider the thermoelectric effect. That is, we consider the charge current flowing through the system due to the different temperatures in the leads. 
Similar to ref.  \cite{PhysRevB.95.075425}, we take a naive but simple approach to determine the stationary charge currents. That is we use the relation between the electrical potential and the amount of charge, which, on the dot, is changed solely by the currents flowing through the tunnel contacts. For that purpose, the phase $\psi$ (corresponding to $\delta V_d$) has to be restored in the action, see App. \ref{app: full dynamics}. However, since we are interested in the stationary currents, we do not need to consider the full quasi-classical dynamics. It is sufficient to consider the retarded AES-like action, eq. \eqref{eq: ret. AESlike action app}, with only the slow retarded kernel function, eq. \eqref{eq: retarded kernel}, that is,
\begin{equation}
i \mathcal{S}_\mathrm{AES}^{R}\! =\! - i\!\! \int\!\! dt\, dt' \sum_{\sigma \sigma'} \mathrm{Im}\! \left[ D_q^{\sigma' \sigma}(t)\, \alpha_{s,\sigma \sigma'}^R(t,t')\, (D_{c}^{\sigma' \sigma}(t'))^* \right]\, .
\end{equation}
Now, the stationary charge currents are obtained by variation with respect to $\psi_q$ and sorting the resulting terms according to the junctions from which they originate. It follows,
\begin{eqnarray}
I_{l \rightarrow dot} & = & I_{d}^l+ I_{h}^l + I_{p}^l\ ,  \\
I_{r \rightarrow dot} & = & I_{d}^r + I_{h}^r\ ,
\end{eqnarray}
where the index $l/r \rightarrow dot$ is for "left-/right-lead to dot" and we defined the pumping current $I_{p}^l = g_l^s \sin^2 \theta_0\, B$, the hybrid charge current $I_{h}^{l/r}= \cos^2 \frac{\theta_0}{2} (I_{h,l/r}^{\uparrow \uparrow}+ I_{h,l/r}^{\downarrow \downarrow}) + \sin^2 \frac{\theta_0}{2} (I_{h,l/r}^{\uparrow \downarrow} + I_{h,l/r}^{\downarrow \uparrow})$, 
and the thermally induced charge current $I_{d}^{l/r}= \cos^2 \frac{\theta_0}{2} (I_{d,l/r}^{\uparrow \uparrow}+ I_{d,l/r}^{\downarrow \downarrow}) + \sin^2 \frac{\theta_0}{2} (I_{d,l/r}^{\uparrow \downarrow} + I_{d,l/r}^{\downarrow \uparrow})$; The hybrid contributions are given by $I_{h,l}^{\sigma \sigma'} = -\rho_\sigma \Gamma_l^{\sigma'}  \frac{\Gamma_\Delta}{\Gamma_\sigma(\theta_0)} \sin^2 \theta_0 B$ and $I_{h,r}^{\sigma \sigma'} = -\rho_\sigma \Gamma_r \frac{\Gamma_\Delta}{\Gamma_\sigma(\theta_0)} \sin^2 \theta_0 B$ and the thermal contributions are given by $I_{d,l}^{\sigma \sigma'} = \rho_{\sigma}' \Gamma_l^{\sigma'} \frac{\Gamma_r}{\Gamma_\sigma(\theta_0)} d$ and $I_{d,r}^{\sigma \sigma'} = \rho_{\sigma}' \Gamma_r (\frac{\Gamma_r}{\Gamma_\sigma(\theta_0)} - 1)\, d$. Explicitly, the currents are given by,
\begin{eqnarray}
I_{d}^{l/r}\!\! &=&\!\! \mp \frac{ \Gamma_r d}{\Gamma_\uparrow(\theta_0) \Gamma_\downarrow(\theta_0)} \big[  \Gamma_r (\rho_\Sigma' \Gamma_\Sigma - \rho_\Delta' \Gamma_\Delta \cos \theta_0) - \nonumber \\
& & \hspace{5em} - \rho_\Sigma'(\Gamma_\Sigma^2- \Gamma_\Delta^2 \cos^2 \theta_0) \big]\ , \\
I_{h}^l\!\! &=&\!\! \frac{-\Gamma_\Delta \sin^2 \theta_0\, B}{\Gamma_\uparrow(\theta_0) \Gamma_\downarrow(\theta_0)} \big[ (\Gamma_\Sigma - \Gamma_r) (\rho_\Sigma \Gamma_\Sigma - \rho_\Delta \Gamma_\Delta \cos \theta_0) + \nonumber \\ 
& & \hspace{3em}+ \Gamma_\Delta \cos \theta_0 ( \rho_\Delta \Gamma_\Sigma - \rho_\Sigma \Gamma_\Delta \cos \theta_0) \big] \ ,\\
I_{p}^l\!\! &=&\!\! \rho_\Sigma \Gamma_\Delta \sin^2 \theta_0\, B\ , \\
I_{h}^r\!\! &=&\!\! \frac{-\Gamma_\Delta \sin^2 \theta_0\, B}{\Gamma_\uparrow(\theta_0) \Gamma_\downarrow(\theta_0)} \left[ \Gamma_r (\rho_\Sigma \Gamma_\Sigma - \rho_\Delta \Gamma_\Delta \cos \theta_0) \right]\ .
\end{eqnarray}
The precession rate of the magnetization, thereby also the external magnetic field, enters the currents twice. First, via its effects on the details of the slow distribution function $F_{s}^\sigma$, giving rise to the hybrid currents $I_{h}^l$ and $I_{h}^r$. Second, via its dynamics\footnote{Note that, this dynamic contribution arises independent of the details of the distribution function. It would also be present and of the same form, if the dot would have an equilibrium distribution.}, it directly gives rise to the pumping current $I_p^l$. This dynamic contribution does not arise for the right contact, because of the spin-independence of $\Gamma_r$.

\begin{figure}
\begin{center}
\includegraphics[width=0.45\textwidth]{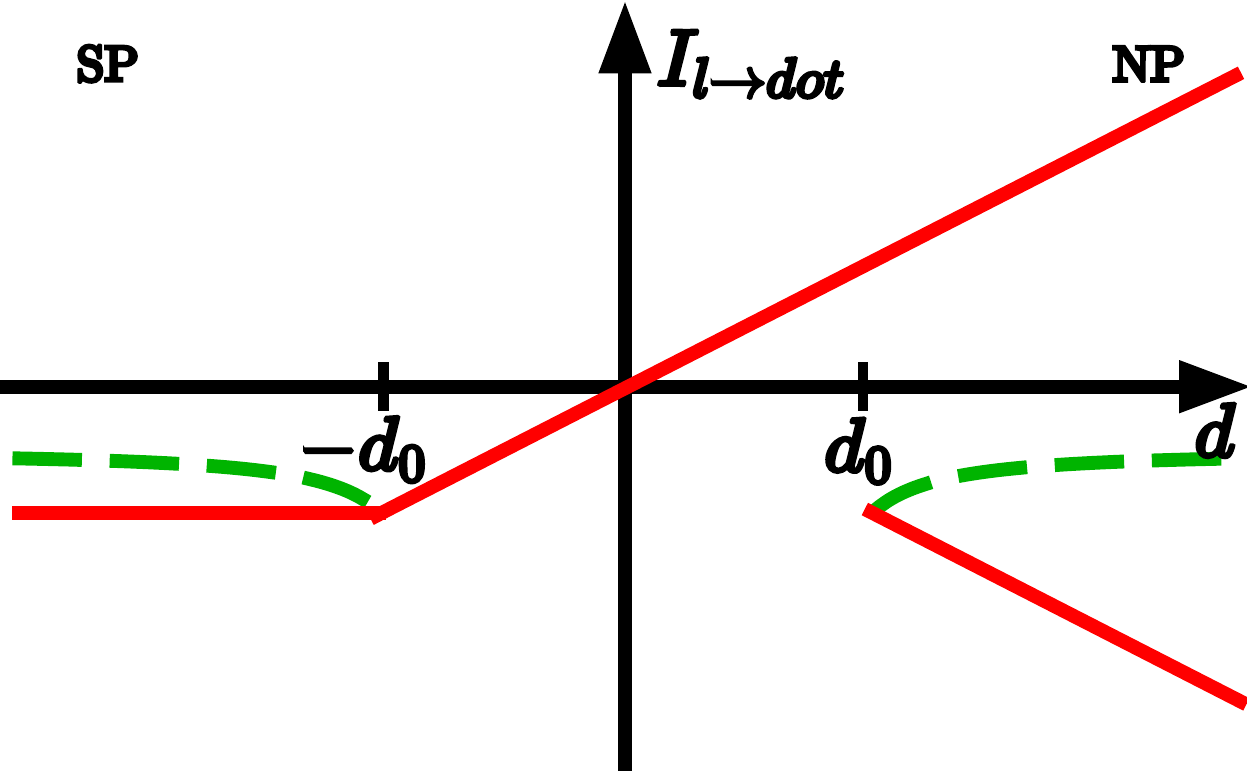}
\end{center}
\caption{For $d= \frac{\pi^2}{3} (T_l^2 - T_r^2)$ and $\Gamma_\Delta<0$, $\rho_\Delta'<0$ and a symmetric density of states, i.e. $\rho_\Delta =0$, $\rho_\Sigma'=0$, we show the charge current $I_{l\rightarrow dot}$ for the stable (red solid) and unstable (blue dotted) stationary solutions of $\cos \theta_0$. Furthermore, we show a hypothetical situation (green dashed), in which the magnetization of the dot makes the angle $\theta_0$ with the $z$-axis, but does not precess.
The value of $\cos \theta_0$ is the same as in the state of persistent precessions at driving $d$. 
In the hypothetical situation the pumping and the hybrid currents are absent.
At $d<-d_0$, i.e., in the regime of stable persistent precessions, we observe a very interesting effect: While the absolute value of the charge current is reduced in comparison to the stationary solution at the north-pole, it is larger than the current for the hypothetical situation without precessions; Thus, we conclude that the precession of the magnetization enhances the thermoelectric effect. For $d> d_0$, we observe a regime of double-stability and the direction of the thermoelectric charge current depends on the orientation of the magnetization.} \label{fig: currents}
\end{figure}

It is straightforward to show that the stationary charge currents balance each other, i.e. $I_{l \rightarrow dot} = - I_{r \rightarrow dot}$. This, of course, must be true for a stationary situation. Interestingly, this balance also holds separately for the "thermally induced" part of the currents $I_{d}^l = - I_{d}^r$ as well as for the hybrid-/pumping-current contributions $I_{h}^l + I_{p}^l = - I_{h}^r$. This splitting might seem superficial at first, since the persistent precession is maintained by the difference in temperatures of the leads. However, for a fixed magnetization in the dot, we expect $I_{h}^l + I_{p}^l$ and $I_{h}^r$ to disappear, whereas $I_{d}^{l/r}$ would remain unchanged. So this splitting also suggests to say that $I_d^{l}$, resp. $I_d^{r}$, describes the standard thermoelectric effect (single-particle), whereas $I_{h}^l + I_{p}^l$, resp. $I_{h}^r$, describe the hybrid-/pumping-part of the thermoelectric effect which is due to the precession of the magnetization (collective). Explicitly, it follows for the stationary charge current,
\begin{eqnarray}
I_{l \rightarrow d}\!\! &=&\!\! \frac{- \Gamma_r d}{\Gamma_\uparrow(\theta_0) \Gamma_\downarrow(\theta_0)} \big[  \Gamma_r (\rho_\Sigma' \Gamma_\Sigma - \rho_\Delta' \Gamma_\Delta \cos \theta_0) - \nonumber \\
& & \hspace{5.5em} - \rho_\Sigma'(\Gamma_\Sigma^2- \Gamma_\Delta^2 \cos^2 \theta_0) \big] +\nonumber \\
& &\!\! + \frac{\Gamma_r \Gamma_\Delta \sin^2 \theta_0\, B}{\Gamma_\uparrow(\theta_0) \Gamma_\downarrow(\theta_0)} (\rho_\Sigma \Gamma_\Sigma - \rho_\Delta \Gamma_\Delta \cos \theta_0)\ ,
\end{eqnarray}
where the term $\propto B$ describes the "hybrid-/pumping-" enhancement of the thermoelectric effect. A dynamically rotating magnetization can be viewed as an adiabatic pump \cite{tserkovnyak2002enhanced}. In this respect, the small magnet can be seen as a thermally driven adiabatic pump. It is physically interesting and may become technically relevant that this pumping effect can be used to enhance the (single-particle) thermoelectric effect.
This is demonstrated for a simple density of states, i.e. for $\rho_\Delta = 0$ and $\rho_\Sigma'=0$ the current is shown in Fig. \ref{fig: currents}.

\section{Summary and Discussion}
We have considered a simple model for a small ferromagnet that can be driven by a thermally induced spin-transfer-torque current. While earlier studies have focused on two lead setups (F$|$I$|$F), we considered a situations with a small ferromagnet between two leads (F$|$I$|$F$|$I$|$N). We have derived the quasi-classical equations of motion for the magnetization dynamics, where the dynamical adjustments of the distribution function to the magnetization are taken into account self-consistently. For that purpose, we extended the approach of ref. \cite{PhysRevB.95.075425} to allow for a simplified treatment of slow coordinates. 

As a result, we obtained the Landau-Lifshitz-Gilbert equation supplemented by a spin-transfer-torque term of the Slonczewski form with two contributions: a thermally induced STT-current $I_d^s(\theta)$ and the dynamically induced hybrid STT-current $I_h^s(\theta, \dot \phi)$. While the hybrid STT-current essentially renormalizes Gilbert-damping, the thermally induced STT-current can be used to drive the magnetization out of its energetic minimum (parallel to the external magnetic field). Furthermore, we determined the stationary charge current corresponding to persistent precessions, and observed again a splitting into two contributions: a single-particle thermoelectric current $I_d^{l}$ (resp. $I_d^r$) and a (collective) hybrid-/pumping-current contribution $I_{h}^l + I_{p}^l$ (resp. $I_{h}^r$) related to the precession of the magnetization. As shown for the simple symmetric density of states, Fig. \ref{fig: stationary} and Fig. \ref{fig: currents}, both current contributions can act in harmony, such that the single-particle thermoelectric current is enhanced by the (collective) pumping current. 

Although the simple model system considered here, may be interesting in its own right, the main purpose of this paper is to provide a basis for further studies on the intersection between mesoscopic physics and spin-(calori-)tronics. 
From this point of view, many options for future work open up. The system should be made more realistic by lifting some of the approximations, most importantly, magnetic anisotropy and internal relaxation mechanism should be included, and the macrospin approximation should be lifted. It would also be interesting to include quantum effects like Coulomb-blockade or zero-bias anomaly. Already for the present simple system, more details could be analyzed, e.g. besides determining the charge current, also heat- and spin-currents should be investigated, and one might want to consider simultaneous thermal and electrical driving. This would be especially relevant for potential technical applications of heat to "useful" energy conversion. Another direction for technical applications would be to search for more adiabatic pumps that could be driven thermally.

\section{Acknowledgements}
We thank S. Backens, Y. Blanter, L. Glazman, M. Kessler, Y. Makhlin, S. Rex, and J. Schmalian for fruitful discussions.
This work was supported by DFG Research Grant SH 81/3-1.
Furthermore, T.L. acknowledges KHYS of KIT and the Feinberg Graduate school of WIS for supporting a stay at WIS;
I.B. acknowledges the Alexander von Humboldt Foundation and the Basic research program of HSE;
Y.G. acknowledges the IMOS Israel-Russia program and the Italia-Israel QUANTRA.

\appendix

\section{Full dynamics\label{app: full dynamics}}
In this appendix, we consider the dynamics of the magnetization length $\eta$ (corresponding to $\delta M$) and the electrical potential $\psi$ (corresponding to $\delta V_d$) in addition to the slow dynamics of $\theta, \dot \phi$. 
It is especially interesting because the relaxation of $\delta M$ happens to take place on a similar time-scale as the adjustments of the distribution function. This demands a more careful treatment than for slow or fast coordinates.

In the following, to distinguish between the different coordinates, we refer to $\theta, \phi$ as $SU(2)$-coordinates, since they are related to the $SU(2)$-rotations $R$, whereas we refer to $\eta, \psi$ as $U(1)$-coordinates, since they are related to the $U(1)$-transformations $U$. The $SU(2)$-coordinates, which have been discussed already in the main text, are included in the slow rotation $D_s= R_k$, whereas we proceed on more general grounds for the $U(1)$-coordinates.

\subsection{Additional contributions to the effective action}
There are two contributions arising from the $U(1)$-coordinates that have to be considered. 
First, we have to take into account the corrections to the classical Green's function, eq. \eqref{eq: expansion of GF},
\begin{equation}
G_c = G_s + G_u\ ,
\end{equation}
with the corrections from $U(1)$-coordinates (u),
\begin{equation}
G_u = G_s (D_k^\dagger \Sigma D_k - R_k^\dagger \Sigma R_k) G_s + ...\ ,
\end{equation}
where we used $D_s = R_k$.
Second, we have to restore the $U(1)$ coordinates in all contributions of the action.

Keeping $\delta M$ and $\delta V_d$, we also have to take into account the zero-mode contributions to the effective action, eqs. \eqref{eq: ZMM} and \eqref{eq: ZMV}.
Terms proportional to the zero-modes $\delta M_0^q$ and $\delta V_{d0}^q$ also appear in
the HS-part of the action which is,
\begin{eqnarray}
i \mathcal S_\mathrm{HS}\!\! & = & \!\! - i \frac{M_0}{2 J} \delta M_q(\omega=0) - i \frac{B}{J} \int dt M_c \sin \theta_c \sin \frac{\theta_q}{2} -\nonumber \\
& &\!\! \hspace{-2em} - \frac{i}{2J} \int dt\, \delta M_c \delta M_q + i \frac{B}{2J} \int dt\, \delta M_q \cos \theta_c \cos \frac{\theta_q}{2} + \nonumber \\
& &\!\! \hspace{-2em} + i (C V_{d0} + N_0) \delta V_{d}^q(\omega=0) + i C \int dt\, \delta V_d^c \delta V_d^q\ , \label{eq: HS app}
\end{eqnarray}
where we dropped constant terms $\propto M_0^2, B^2, V_{d0}^2$.

For the WZNW-contribution, the sole change is in the length of the spin $S$, eq. \eqref{eq: slow spin}. In the equations of motion, these fluctuations would lead to the corrections of order $1/S$, which we disregard.
Justified by the large value of $S$, we also disregard the LZ-contribution to the effective action.

The most important changes are in the AES-like contribution. Restoring $\eta$ and $\psi$, the full gauge transformation $D_c, D_q$ will appear in the retarded part,
\begin{equation}
i \mathcal{S}_\mathrm{AES}^{R}\! =\! - i\!\! \int\!\! dt\, dt' \sum_{\sigma \sigma'} \mathrm{Im}\! \left[ D_q^{\sigma' \sigma}(t)\, \alpha_{\sigma \sigma'}^R(t,t')\, (D_{c}^{\sigma' \sigma}(t'))^* \right]\ . \label{eq: ret. AESlike action app}
\end{equation}
Furthermore, the retarded kernel function now becomes,
\begin{eqnarray}
\alpha_{\sigma \sigma'}^R(t,t')\! & = &\! \mathrm{tr}\! \left[G_{ \sigma}^R(t,t') \Sigma^K_{\sigma'}(t'-t) + G_{ \sigma}^K(t,t') \Sigma^A_{\sigma'}(t'-t) \right]\nonumber \\ 
& = & \alpha_{s,\sigma \sigma'}^R(t,t') + \alpha_{u,\sigma \sigma'}^R(t,t')\ ,  \label{eq: retarded kernel app}
\end{eqnarray}
where the slow contribution is known from the main text, eq. \eqref{eq: retarded kernel}. The new contribution arising from $U(1)$-coordinates is given by,
\begin{eqnarray}
& &\hspace{-1.5em} \alpha_{u,\sigma \sigma'}^R(t,t''') = \nonumber \\
& &\hspace{-1.5em}  = \mathrm{tr}\! \left[G_{u \sigma}^R(t,t''') \Sigma^K_{\sigma'}(t'''-t) + G_{u \sigma}^K(t,t''') \Sigma^A_{\sigma'}(t'''-t) \right]\nonumber \\
& &\hspace{-1.5em}  = \int dt'\, dt'' (U_{k\sigma}^\dagger(t') U_{k\sigma}(t'') - 1) \beta^{R}_{\sigma \sigma'}(t,t',t'',t''')\ , \label{eq: alpha def}
\end{eqnarray}
where we used $D_k= R_k U_k$ with $U_k = U_c|_{q=0}$ and the $SU(2)$-rotations $R_k$ are absorbed into,
\begin{eqnarray}
& & \beta^{R}_{\sigma \sigma'}(t,t',t'',t''') = \nonumber  \\
& &  =\!\! \mathrm{tr} \!\bigg[ G_{s\sigma}^R(t,t')\! \left[ R_k^\dagger \Sigma^R R_k \right]_{\sigma \sigma}\!(t',t'') G_{s\sigma}^R(t'',t''') \Sigma_{\sigma'}^{K}(t'''\!-\!t)\!\! +\nonumber \\
& & + G_{s\sigma}^K(t,t')\! \left[ R_k^\dagger \Sigma^A R_k \right]_{\sigma \sigma}\!(t',t'') G_{s\sigma}^A(t'',t''') \Sigma_{\sigma'}^{A}(t'''\!-\!t) +\nonumber \\
& &  + G_{s\sigma}^R(t,t')\! \left[ R_k^\dagger \Sigma^K R_k \right]_{\sigma \sigma}\!(t',t'') G_{s\sigma}^A(t'',t''') \Sigma_{\sigma'}^{A}(t'''\!-\!t) +\nonumber \\
& & + G_{s\sigma}^R(t,t')\! \left[ R_k^\dagger \Sigma^R R_k \right]_{\sigma \sigma}\!(t',t'') G_{s\sigma}^K(t'',t''') \Sigma_{\sigma'}^{A}(t'''\!-\!t) \bigg]\ .\nonumber \\
 \label{eq: beta kernel}
\end{eqnarray}
The calculation of the retarded kernel function $\alpha_{u,\sigma \sigma'}^R(t,t''')$ is not trivial but it is also not really illuminating, thus we shift it to the end of this appendix \ref{sec: calc of non-slow kernel}. Using the slowness of $\theta$ and $\dot \phi$ and disregarding terms of $\mathcal{O} \left( \frac{1}{S} \right)$, we obtain,
\begin{eqnarray}
\alpha_{u,\sigma \sigma'}^R(t,t''')\!\! & = &\!\! i 2 g_{\sigma \sigma'} \delta(t-t''') \Gamma_\sigma(\theta(t)) \times \nonumber \\
& &\!\! \times \int_{-\infty}^t\!\!\! dt'\, e^{-2 \Gamma_\sigma(\theta(t)) (t-t')} \, U_{k\sigma}^*(t') \dot U_{k\sigma}(t')\ .\nonumber \\ \label{eq: alpha result}
\end{eqnarray}
It is now straightforward to insert this kernel-function back into the AES-like action, eq. \eqref{eq: ret. AESlike action app}. Then a variation with respect to quantum components yields the quasi-classical equations of motion.

\subsection{Quasiclassical equations of motion}
We add up all contributions to the effective action and, then, expand to first order in quantum components $\theta_q, \phi_q, \eta_q, \psi_q$. Afterwards the variation with respect to quantum components is trivial and we obtain the coupled equations of motion,
\begin{eqnarray}
\sin \theta\, \dot \phi\!\! & = &\!\! - \sin \theta\, B\ , \label{eq: eom phi int} \\
\sin \theta\, \dot \theta\!\! & = &\!\! \frac{\sin^2 \theta}{S} \Bigg\lbrace \bigg[ \tilde g(\theta) \dot \phi  - I_{h}^s(\theta)- I_D^s(\theta) \bigg] +\nonumber \\ 
& &\hspace{3em} +\bigg[ \Gamma_\Delta \sum_\sigma \rho_\sigma \Big( \delta V_d - 2 \Gamma_\sigma(\theta) R_V^\sigma \Big) \bigg] - \nonumber \\
& &\hspace{3em} -\bigg[ \frac{\Gamma_\Delta}{2} \sum_\sigma \sigma \rho_\sigma \Big( \delta M - 2 \Gamma_\sigma(\theta) R_M^\sigma \Big)\bigg]\Bigg\rbrace\ , \nonumber \\ \label{eq: eom theta int}\\
\frac{1}{J} \delta \dot M\!\! & = &\!\! + \sum_\sigma \rho_\sigma \Gamma_\sigma (\theta) \Big( \delta M - 2 \Gamma_\sigma(\theta) R_M^\sigma \Big) - \nonumber \\ 
& &\!\! - \sum_\sigma \sigma 2 \rho_\sigma \Gamma_\sigma(\theta) \Big( \delta V_d - 2 \Gamma_\sigma(\theta) R_V^\sigma \Big)\ , \label{eq: eom M int}\\
C \delta \dot V_d\!\! & = & \!\! - \sum_\sigma 2 \rho_\sigma \Gamma_\sigma(\theta) \Big( \delta V_d - 2 \Gamma_\sigma(\theta) R_V^\sigma \Big) + \nonumber \\
&  &\!\! + \sum_\sigma \sigma \rho_\sigma \Gamma_\sigma(\theta) \Big( \delta M - 2 \Gamma_\sigma(\theta) R_M^\sigma \Big)\ ,\label{eq: eom V int}
\end{eqnarray}
where we resubstituted $\dot \eta_c = \delta M_c$ and $\dot \psi_c = \delta V_d^c$ and only leading order terms in $1/S$ were kept. Furthermore, we introduced the retarded integrals,
\begin{eqnarray}
R_V^\sigma & = & \int_{-\infty}^t\!\!\! dt'\, e^{-2 \Gamma_\sigma(\theta)(t-t')} \delta V_d(t')\ , \\
R_M^\sigma & = & \int_{-\infty}^t\!\!\! dt'\, e^{-2 \Gamma_\sigma(\theta)(t-t')} \delta M(t')\ .
\end{eqnarray}
The method described above will usually lead to equations of motion of the integro-differential-type. The retarded integrals $R_V^\sigma$ and $R_M^\sigma$ originate from the kernel $\alpha_{u,\sigma \sigma'}^{R}(t, t''')$ which arise from the corrections for $U(1)$-coordinates. We think that the physical origin of this retardation effect is that the distribution function for spin $\sigma$ changes on the time-scale determined by the inverse level broadening $1/\Gamma_\sigma(\theta_0)$. On those time-scales, the information about past values of the coordinates is stored in the dynamic distribution function. Would $\delta V_d$ and $\delta M$ be slow (approx. constant) on this time-scale, then the integrals could be easily performed and the retardation effect would be gone. However, $\delta V_d$ is fast compared to the distribution function and $\delta M$ changes typically on roughly the same time-scale as the distribution function. Therefore, we cannot assume them to be slow and, in turn, we should carefully consider $R_V^\sigma$ and $R_M^\sigma$.

By making use of the Fourier-transformation, we can recast the integro-differential equations \eqref{eq: eom M int} and \eqref{eq: eom V int} into differential equations \eqref{eq: eom M diff} and \eqref{eq: eom V diff}. Thereby, we assume $\theta$ to be approximately constant, which means to disregard corrections of higher order in $1/S$. Similarly, the second and third line of equation \eqref{eq: eom theta int} is recasted into the second line of equation \eqref{eq: eom theta diff}.
\begin{eqnarray}
& & \hspace{-0em} \sin \theta\, \dot \phi = - \sin \theta\, B\ , \label{eq: eom phi int} \\
& & \hspace{-0em} \sin \theta\, \dot \theta = \frac{\sin^2 \theta}{S} \bigg[ \tilde g(\theta) \dot \phi  - I_{h}^s(\theta)- I_D^s(\theta) + \nonumber \\ 
& & \hspace{6.5em} +\Gamma_\Delta (\rho_\Sigma + C)\, \delta V_d - \frac{\rho_\Delta \Gamma_\Delta}{2}\, \delta M \bigg]\ , \label{eq: eom theta diff} \\
& & \hspace{-0em} \delta \dot M= + \left[ \frac{g_\uparrow(\theta)}{2} \left(J-\frac{1}{\rho_\uparrow}\right) + \frac{g_\downarrow(\theta)}{2} \left(J-\frac{1}{\rho_\downarrow}\right) \right] \delta M - \nonumber \\ 
& &\hspace{2.9em} - \left[ g_\uparrow(\theta) \left(1+\frac{C}{2 \rho_\uparrow}\right) - g_\downarrow(\theta) \left(1+\frac{C}{2 \rho_\downarrow}\right) \right] J\, \delta V_d\ , \nonumber \\ \label{eq: eom M diff} \\
& & \hspace{-0em} \delta \dot V_d = - \left[g_\uparrow(\theta) \left(\frac{1}{C} + \frac{1}{2 \rho_\uparrow} \right) + g_\downarrow(\theta) \left( \frac{1}{C} + \frac{1}{2 \rho_\downarrow}\right) \right] \delta V_d +\nonumber \\
& &\hspace{2.9em} + \left[g_\uparrow(\theta) \left( 1 - \frac{1}{\rho_\uparrow J}\right)  - g_\downarrow(\theta) \left( 1- \frac{1}{\rho_\downarrow J} \right) \right]\!\! \frac{1}{2 C} \delta M\ , \nonumber \\ \label{eq: eom V diff}
\end{eqnarray}
where we defined $g_\sigma(\theta)= 2 \rho_\sigma \Gamma_\sigma(\theta)$. We note that the term $\propto C$ in eq. \eqref{eq: eom theta diff} and all terms that explicitly contain $\frac{1}{\rho_\sigma}$ originate from the correction to the Green's function $G_u$, due to the $U(1)$-coordinates.

To gain a deeper insight into the physics of those contributions arising from $G_u$, we consider the simple case with $\rho_\uparrow = \rho_\downarrow = \rho$ (e.g. for symmetric density of states) and $\Gamma_l^\uparrow = \Gamma_l^\downarrow$ (e.g. both leads non-magnetic). Then, the equations of motion for $\delta \dot M$ and $\delta \dot V_d$ decouple and we obtain,
\begin{eqnarray}
\delta \dot M\!\! & = &\!\! g \left( J - \frac{1}{\rho} \right) \delta M\ , \\ 
\delta \dot V_d\!\! & = &\!\! - 2g \left( \frac{1}{C} + \frac{1}{2\rho} \right) \delta V_d\ ,
\end{eqnarray}
where we defined $g= 2 \rho \Gamma_\Sigma$.

The equation for $\delta M$ is easy to understand. The exchange interaction $\propto J$ tends to align spins on the dot and thus tries to increase the magnetization. If there was no competing effect, the magnetization on the dot would grow without bounds by acquiring more and more electrons with their spins in parallel. However, the Pauli-exclusion principle forbids two electrons to occupy the same state and thus for each spin that is added to the dot a higher level (level spacing $\frac{1}{\rho}$) has to be occupied by an electron, i.e. more energy has to be paid. The dynamics of $\delta M$ is described by the competition of both effects. Note that fluctuations $\delta M$ should always relax to zero, since otherwise we would not have chosen the correct $M_0$. And indeed it is $\frac{1}{\rho} > J$ in the Stoner-regime after a magnetization has been built up on the dot\footnote{Note that $\rho$ is the density of states at the shifted Fermi-energy $\mu \mp \frac{M_0}{2} + V_{d0}$}. So, we find that the term $\frac{1}{\rho}$ is essential for the dynamics of $\delta M$. Tracing back the origin of $\frac{1}{\rho}$, we find this term to arise from the Keldysh part of $G_u$, i.e. the contribution $U(1)$-coordinates; it is, thus, related to the dyncamic change in the distribution function with fluctuations of $\delta M$. While this might be clear from the point of view of the Stoner-transition physics, it is also interesting to view this from a more formal perspective. The dynamics of $\delta M$ takes place roughly at the same time-scale as the change in distribution function. Thus, the interplay of $\delta M$ with the distribution function can (and turned out to) be important for its dynamics.

The situation for $\delta V_d$ is analog but simpler. Instead of the attractive exchange interaction, there is repulsive Coulomb interaction $\propto \frac{1}{C}$. Thus, Pauli-exclusion assists Coulomb interaction instead of competing with it. The equation for $\delta V_d$ describes the standard charge relaxation through a resistor if the (effective) electrochemical potential is not at its stationary value. The capacity contribution of $\frac{1}{C}$ is related to the change of the electrochemical potential by addition of charges, i.e. the change in electrical potential. The contribution of $\frac{1}{\rho}$ is related to the change of the electrochemical potential by addition of particles, i.e. the change in chemical potential; it is also known as quantum capacity. From a formal point of view, we note that the relaxation of $\delta V_d$ is much faster than the time-scale of changes in the distribution, i.e. the distribution function has not enough time to react to changes of $\delta V_d$. Thus, the correction to the Coulomb repulsion should be quite small. This is indeed the case: For systems that are large compared to the atomic scale the quantum capacity is a small correction, i.e. $\frac{\rho}{C} \gg 1$.

\subsection{Zero-mode equations \label{sec: zero modes}}
We emphasize that the equations of motion do not determine the stationary values $M_0$ and $V_{d0}$. To fix those values, we have to consider the contributions from the quantum zero-mode effective actions, eqs. \eqref{eq: ZMM} and \eqref{eq: ZMV} in combination with the zero-mode parts from the HS-part, eq. \eqref{eq: HS app}. Variation with respect to the quantum zero-modes $\delta M^q_0$ and $\delta V_{d0}^q$ yields\footnote{Whereas a simple-minded variation 
would produce here $G_c^K(t,t)/2$ instead of $G_c^<(t,t)$, a proper regularization of the same time expressions,
see chapter 2.8 in ref.~\cite{KamenevBook} and ref.~\cite{doi:10.1080/00018730902850504} leads to stated results.},
\begin{eqnarray}
 &&\frac{M_0}{2 J} =-\frac{i}{2} \frac{1}{2 T_K}\int_{-T_K}^{T_K}\! dt\, \mathrm{tr}\left[ G_c^<(t,t) \sigma_z \right] \ , \\
 &&C V_{d0}    = \frac{1}{2 T_K} \int_{-T_K}^{T_K}\! dt\, \left(- i \,\mathrm{tr}\left[ G_c^<(t,t) \right]-N_0\right)\ . 
 \nonumber\\
\end{eqnarray}
The first equation can be read in two related ways: On one hand this relates the magnetization $M_0$ to the (time-average of the) spin $S(t)$ by $M_0 = 2J \langle S \rangle$; on the other hand $S(t)$ depends on the Green's function, which depends on $M_0$ and, thus, it can be read as the self-consistency equation for the magnetization length $M_0$. The second equation is the analog for the electrical potential $V_{d0}$ with the charge $Q(t)= -i\, \mathrm{tr}\left[ G_c^<(t,t) \right]-N_0$. The stationary values $M_0$ and $V_{d0}$ can be determined from these (coupled) self-consistency equations.

\subsection{Calculation of the $U(1)$-correction to the retarded kernel function \label{sec: calc of non-slow kernel}}
Note that only the third term in $\beta^{R}$ contributes to the action. The other three terms drop out, since the factor $(U_{k\sigma}^\dagger(t') U_{k\sigma}(t'') - 1)$ vanishes in combination with the time-local self-energies $\Sigma^{R/A}(t'-t'') \propto \delta(t'-t'')$. In the following, we only keep the third term for which we find,
\begin{eqnarray}
& &\hspace{-0em}\beta^{R}_{\sigma \sigma'}(t,t'\!,t''\!\!,t''')\!\! =\!\! \!\! \int\!\!\! \frac{d\omega_1}{2\pi}\!\!\! \int\!\!\! \frac{d\omega_2}{2\pi}\!\!\! \int\!\!\! \frac{d\omega_3}{2\pi}\!\!\! \int\!\!\! \frac{d\omega'}{2\pi} e^{-i[\omega_1 t_1+\omega_2 t_2 +\omega_3 t_3]}\! \times \nonumber  \\
& &\hspace{-0em}\, \times \mathrm{tr} \bigg[G_{s\sigma}^R(\cmt_1, \omega_1+ \omega')\! \left[ R_k^\dagger \Sigma^K R_k \right]_{\sigma \sigma}\!(\cmt_2, \omega_2 + \omega') \times \nonumber \\
& &\hspace{2em}\, \times G_{s\sigma}^A(\cmt_3, \omega_3+\omega')\, \Sigma_{\sigma'}^{A}(\omega')\bigg]\ ,
\end{eqnarray}
where we have written $\cmt_1 = \frac{t+t'}{2}, t_1=(t-t')$ and $\cmt_2 = \frac{t'+t''}{2}, t_2=(t'-t'')$ and $\cmt_3 = \frac{t''+t'''}{2}, t_3=(t''-t''')$ for brevity. Insertion of the slow Green's function and slowly rotated self-energy yields,
\begin{eqnarray}
& &\hspace{-0em}\beta^{R}_{\sigma \sigma'}(t,t'\!,t''\!\!,t''')\!\! =\!\! \!\! \int\!\!\! \frac{d\omega_1}{2\pi}\!\!\! \int\!\!\! \frac{d\omega_2}{2\pi}\!\!\! \int\!\!\! \frac{d\omega_3}{2\pi}\!\!\! \int\!\!\! \frac{d\omega'}{2\pi} e^{-i[\omega_1 t_1+\omega_2 t_2 +\omega_3 t_3]}\! \times \nonumber  \\
& &\hspace{-0em}\, \times \frac{2 \Gamma_\sigma(\theta(\cmt_2)) (\Gamma_l^{\sigma'} + \Gamma_r)}{[\omega' + \omega_1 + i \Gamma_\sigma(\theta(\cmt_1))] [\omega' + \omega_3 - i \Gamma_\sigma(\theta(\cmt_3))]} \times \nonumber \\ 
& &\hspace{-0em} \times\!\! \sum_\alpha\! \bigg[\! \Big(\! F_s^\sigma(\cmt_2,\omega'\!\! +\! \omega_2\!\! +\! \xi_{\alpha \sigma})\! -\! F_s^\sigma(\cmt_2,\omega'\!\! +\! \xi_{\alpha \sigma})\! \Big)\! + \nonumber \\ 
& &\hspace{3em} + F_s^\sigma(\cmt_2,\omega'\!\! +\! \xi_{\alpha \sigma}) \bigg]\ , \label{eq: beta freq}
\end{eqnarray}
where we have shifted the integration over $\omega' \rightarrow \omega' + \xi_{\alpha \sigma}$ and to the slow distribution function $F_s^\sigma(\cmt_2,\omega'\!\! +\! \omega_2\!\! +\! \xi_{\alpha \sigma})$, we subtracted and added the same slow distribution function but with $\omega_2 \rightarrow 0$. Now, we can easily calculate the difference,
\begin{equation}
\sum_\alpha\! \Big(\! F_s^\sigma(\cmt_2,\omega'\! +\! \omega_2\! +\! \xi_{\alpha \sigma})\! -\! F_s^\sigma(\cmt_2,\omega'\! +\! \xi_{\alpha \sigma})\! \Big)\! \approx\! 2 \rho_\sigma \omega_2\ ,
\end{equation}
where corrections\footnote{Note that the shifted density of states $\rho_\sigma$ in this result is unbroadend and thus it is slightly different from the density of states as introduced in the main text. However, we assume the broadening $\Gamma_\sigma(\theta)$ to be much smaller than $M_0$ and $T_{l/r}$. Then, this difference leads to corrections of $\mathcal O (\frac{1}{S})$ which we disregard.} of  $\mathcal O (\frac{1}{S})$ are disregarded and only values of $\omega' \ll M_0$ are assumed to be relevant. Since the remaining (added) distribution function $F_s^\sigma(\cmt_2,\omega'\!\! +\! \epsilon_\alpha\!\! -\! \frac{M_0}{2} \sigma)$ is independent of $\omega_2$, it would lead to a term in $\beta^R_{\sigma \sigma'}$ that is $\propto \delta (t'-t'')$ and, therefore, it would vanish in combination with the factor $(U_{k\sigma}^\dagger(t') U_{k\sigma}(t'') - 1)$. We drop this term already in $\beta^R_{\sigma \sigma'}$. It is, then, straightforward to perform the integrations over frequencies in eq. \eqref{eq: beta freq} and insert it back into eq. \eqref{eq: alpha def} to obtain the result for the retarded kernel function eq. \eqref{eq: alpha result}.

\section{Approximation for slow coordinates}
In this rather formal appendix, we discuss the approximations for the slowness of the coordinates $\theta, \dot \phi$.

\subsection{Slowly rotated self-energy \label{app: slowly rotated self energy}}
In the main text, the rotated self-energy $D_k^\dagger \Sigma D_k$ is split into a slow part $R_k^\dagger \Sigma R_k$ and the rest $(D_k^\dagger \Sigma D_k - R_k^\dagger \Sigma R_k)$. The slowly rotated self-energy is then given by,
\begin{equation}
(R_k^\dagger \Sigma R_k)_{\sigma \sigma''} (t,t') = \sum_{\sigma'} (R_k^\dagger)_{\sigma \sigma'}(t) \Sigma_{\sigma'}(t-t') R_{k \sigma' \sigma''}(t)\ .
\end{equation}
Its spin-diagonal part is,
\begin{eqnarray}
& &\hspace{-1.5em}(R_k^\dagger \Sigma R_k)_{\sigma \sigma} (t,t') = \nonumber \\
& = & \Sigma_\sigma (t-t') \cos \frac{\theta (t)}{2} \cos \frac{\theta (t')}{2} e^{i \sigma \int_{t'}^{t} dt''\, \dot \phi(t'') \frac{1-\cos \theta(t'')}{2}} +\nonumber \\
& + & \Sigma_{\bar \sigma} (t-t') \sin \frac{\theta(t)}{2} \sin \frac{\theta(t')}{2} e^{i \bar \sigma \int_{t'}^{t} dt''\, \dot \phi(t'') \frac{1+\cos \theta(t'')}{2}}\ . \nonumber \\
\end{eqnarray}

For the retarded and advanced part, we can use the time-locality of the unrotated self-energy $\Sigma_\sigma^{R/A}(t-t') = \mp i (\Gamma_l^\sigma+\Gamma_r) \delta(t-t')$ to obtain,
\begin{equation}
(R_k^\dagger \Sigma^{R/A} R_k)_{\sigma \sigma} (\cmt,\omega) = \mp i \Gamma_\sigma(\theta(\cmt))\ ,
\end{equation}
for the spin-diagonal part, where we introduced the "center of mass"-time $\cmt=\frac{t+t'}{2}$ and the relative time $\tilde t=t-t'$ (the Wigner coordinates) and performed the Fourier transform with respect to $\tilde t$. For the spin-off-diagonal part it follows,
\begin{equation}
(R_k^\dagger \Sigma^{R/A} R_k)_{\sigma \bar \sigma} (\cmt,\omega)\!\! =\!\! \pm i \Gamma_\Delta\sin \theta(\cmt) e^{i \bar \sigma \int^{\cmt}_{-\infty} dt\, \dot \phi(t) \cos \theta(t)}\ .
\end{equation}

For the Keldysh part of the slowly rotated self-energy, the situation is more complicated, since the Keldysh part of the unrotated self-energy $\Sigma_{\sigma}^K(t-t') = - 2i (\Gamma_l^\sigma F_l(t-t') + \Gamma_r F_r(t-t'))$ is not local in time. The typical time-scale of $F_{l/r}(t-t')$ is given by the inverse temperatures of the leads, i.e. $1/T_{l/r}$, which is the correlation-time of thermal noise. Assuming $\theta$ and $\dot \phi$ to be approximately constant on this time-scale, i.e. thermal noise appears to be white, we obtain for the spin-diagonal part,
\begin{eqnarray}
& &\hspace{-2em}(R_k^\dagger \Sigma^K R_k)_{\sigma \sigma} (\cmt,\omega) \approx \nonumber \\
& &\hspace{-0em} \approx \cos^2\! \frac{\theta}{2}\, \Sigma_\sigma(\omega + \sigma \omega_-) + \sin^2\! \frac{\theta}{2}\, \Sigma_{\bar \sigma}(\omega + \sigma \omega_+) =\nonumber \\
& & \hspace{-0em}= - 2i\, \Gamma_\sigma(\theta)\, F_s^\sigma (\cmt,\omega)\ ,
\end{eqnarray}
with $\theta= \theta(\cmt)$ and $\omega_\pm=\dot \phi(\cmt) \frac{1\pm \cos \theta(\cmt)}{2}$. For the spin-off-diagonal parts, we obtain,
\begin{eqnarray}
& &\hspace{-1em}(R_k^\dagger \Sigma^K R_k)_{\sigma \bar \sigma} (\cmt,\omega) \approx - \sigma \frac{\sin \theta(\cmt)}{2} e^{i \bar \sigma \int^{\cmt}_{-\infty} dt\, \dot \phi(t) \cos \theta(t)} \times \nonumber \\
& &\hspace{1em} \times  \left[ \Sigma_\sigma^K \Big( \omega + \sigma \frac{\dot \phi(\cmt)}{2} \Big) - \Sigma_{\bar \sigma}^K \Big( \omega + \bar \sigma \frac{\dot \phi(\cmt)}{2} \Big) \right]  \ .
\end{eqnarray}
We note that the spin-diagonal contributions depend on time only through the slow coordinates, i.e. $\theta$ and $\dot \phi$. In contrast, the spin-off-diagonal contributions include a phase-factor, which can change fast. The phase depends on time roughly like $\cos \theta(\cmt)\, B\,\cmt$, i.e. it is of intermediate speed or even fast, if $B$ is larger than the level broadening. Therefore, the spin-off-diagonal contributions should not have been included into the slowly rotated self-energy. However, due to the large magnetization, we are going to disregard spin-off-diagonal contributions anyway.

Next, we consider the gradient expansion, which is essential to determine the slow Green's function. Afterwards, we determine the slow Green's function and, thereby, obtain another criterion that must be satisfied by $\theta$ and $\dot \phi$ to pass as slow coordinates.

\subsection{Gradient expansion \label{app: gradient exp new}}
The gradient expansion for the convolution of two functions $f(t,t'') = \int dt' g(t,t') h(t',t'')$, is easily found in literature, e.g. \cite{KamenevBook, altland2010condensed}. Following those ideas, we give a short schematic derivation which is tailor-made for extension to the case of three functions $f(t,t''') = \int dt' \int dt'' g(t,t') h(t',t'') k(t'',t''')$.

At first, we change to "center of mass"-time and "relative"-time for all functions, i.e. $\tilde f(\frac{t+t''}{2},t-t'') = f(t,t'')$, $\tilde g(\frac{t+t'}{2},t-t') = g(t,t')$, and $\tilde h(\frac{t'+t''}{2},t'-t'') = h(t',t'')$ is introduced, where the $\tilde{\ \ }\ $-notation is introduced to formally distinguish between different arrangements of time-arguments. We define $\cmt=\frac{t+t''}{2}$ and $\tilde t = t-t''$ and use the Fourier-transformations in time-differences to obtain,
\begin{eqnarray}
\tilde f(\cmt,\omega)\!\! & = &\!\!\! \int\!\!\! d\tilde t\!\! \int\!\!\! d t'\!\! \int\!\!\!\frac{d\omega'}{2\pi}\!\!\!  \int\!\!\! \frac{d\omega''}{2\pi} e^{i [\omega \tilde t- \omega' (\cmt+\frac{\tilde t}{2}-t')- \omega'' (t'- \cmt +\frac{\tilde t}{2})]}\times \nonumber \\ 
& & \hspace{-0.5em} \times \tilde g \left(\frac{\cmt +\frac{\tilde t}{2}-t'}{2},\omega'\right)\, \tilde h \left(\frac{t'+\cmt -\frac{\tilde t}{2}}{2},\omega''\right)\, .
\end{eqnarray}
Being guided by the desired zeroth order result, see eq. \eqref{eq: two gradient formal zeroth order} below, we redefine time- and frequency-integration variables to obtain,
\begin{eqnarray}
& &\hspace{-3.2em}\tilde f(\cmt, \omega)\! = \!\!\! \int\!\!\! dt_1\!\!\! \int\!\!\! dt_2\!\!\! \int\!\!\! \frac{d\omega_1}{2\pi}\!\!\! \int\!\!\! \frac{d\omega_2}{2\pi}\, e^{-i(\omega_1 t_1+\omega_2 t_2)}\!\!\times \nonumber \\
& &\ \times \tilde g \bigg(\cmt + \frac{t_2}{2},\omega + \omega_1\bigg)\, \tilde h \bigg(\cmt -\frac{t_1}{2}, \omega + \omega_2\bigg)\ ,
\end{eqnarray}
such that the functions on the right side have the form $\tilde g( \cmt + ...\, , \omega + \omega_1)$ and $\tilde h(\cmt + ...\, , \omega + \omega_2)$. The idea is now to formally expand $\tilde g$ in $\omega_1$ and $\tilde h$ in $\omega_2$ and integrate the resulting series term-wise. At first the integrals over $\omega_1, \omega_2$ are performed, leading to derivatives of $\delta$-functions. Then the integration over times $t_1, t_2$, can be performed using partial integration. 
The result of this procedure can be written in a quite compact form,
\begin{equation}
\tilde f(\cmt, \omega) = \mathrm{exp}\left[{-\frac{i}{2} (\partial_{\cmt}^{\bar h} \partial_\omega^g + \partial_{\cmt}^{g} \partial_\omega^h)}\right] \tilde g(\cmt, \omega) \tilde h(\cmt,\omega)\ ,\label{eq: gradient for two}
\end{equation}
where, as usual, subscripts indicate which variable to differentiate. Superscripts indicate on which function the derivative is applied. A bar in the superscript indicates to include a factor of $(-1)$.
Keeping only the zeroth order term from the exponential we obtain,
\begin{equation}
\tilde f_0(\cmt,\omega) = \tilde g(\cmt,\omega)\, \tilde h(\cmt,\omega)\ ,\label{eq: two gradient formal zeroth order}
\end{equation}
while,  for example,  the first order term is given by $\tilde f_1(\cmt,\omega)=-\frac{i}{2} (\partial_{\cmt}^{\bar h} \partial_\omega^g + \partial_{\cmt}^{g} \partial_\omega^h)\, \tilde g(\cmt, \omega) \tilde h(\cmt,\omega)= - \frac{i}{2} \left\lbrace \left[ \partial_\omega \tilde g(\cmt,\omega) \right] \left[ - \partial_{\cmt} \tilde h(\cmt,\omega) \right] + \left[ \partial_{\cmt} \tilde g(\cmt,\omega) \right] \left[ \partial_\omega \tilde h(\cmt,\omega) \right] \right\rbrace$.

It is now straightforward to extend these ideas to three functions $f(t,t''') = \int dt' \int dt'' g(t,t') h(t',t'') k(t'',t''')$. As intermediate result, before expansion, we obtain,
\begin{eqnarray}
& &\hspace{-0.3em}\tilde f(\cmt, \omega)\! = \!\!\! \int\!\!\! dt_1\!\!\! \int\!\!\! dt_2\!\!\! \int\!\!\! dt_3\!\!\! \int\!\!\! \frac{d\omega_1}{2\pi}\!\!\! \int\!\!\! \frac{d\omega_2}{2\pi}\!\!\! \int\!\!\! \frac{d\omega_3}{2\pi}\, e^{-i(\omega_1 t_1+\omega_2 t_2 +\omega_3 t_3)}\!\!\times \nonumber \\
& &\ \times \tilde g \bigg(\cmt+ \frac{t_2+t_3}{2},\omega + \omega_1\bigg)\, \tilde h \bigg(\cmt+\frac{t_3-t_1}{2}, \omega + \omega_2\bigg) \times \nonumber \\ & &\ \times \tilde k \bigg(\cmt - \frac{t_1 + t_2}{2},\omega+ \omega_3\bigg)\ .
\end{eqnarray}
Note that the form is again guided by the desired zeroth order result, eq. \eqref{eq: gradient expansion three functions zero}.
After expansion in $\omega_1, \omega_2, \omega_3$, term-wise integration over $\omega_1, \omega_2, \omega_3$, and partial integration of $t_1,t_2,t_3$, we obtain the compact result,
\begin{eqnarray}
\tilde f(\cmt, \omega) & = &  \mathrm{exp}\left[{-\frac{i}{2} (\partial_{\cmt}^{\bar h \bar k} \partial_\omega^g + \partial_{\cmt}^{g \bar k} \partial_\omega^h + \partial_{\cmt}^{g h} \partial_\omega^k )} \right] \times \nonumber 
\\ & &\hspace{1.2em} \times \, \tilde g(\cmt , \omega) \tilde h(\cmt,\omega) \tilde k(\cmt,\omega)\ . \label{eq: gradient expansion three}
\end{eqnarray}
As before, superscripts indicate on which functions a derivative should be applied and a bar indicates to include an additional factor of $(-1)$, e.g. $\partial_{\cmt}^{g \bar k}\, (\tilde g\, \tilde k) = (\partial_{\cmt} \tilde g) \tilde k + \tilde g (-\partial_{\cmt} \tilde k)$ and $\partial_{\cmt}^{\bar h \bar k} (\tilde h\, \tilde k) = (-\partial_{\cmt} \tilde h) \tilde k + \tilde h (-\partial_{\cmt} \tilde k)$. For the zeroth order term it follows,
\begin{equation}
\tilde f_0(\cmt,\omega) = \tilde g(\cmt,\omega)\, \tilde h(\cmt,\omega)\, \tilde k(\cmt,\omega)\ .\label{eq: gradient expansion three functions zero}
\end{equation}
This zeroth order result could probably be guessed right away. The main point of the derivation is to obtain a formal criterion for "slow" dynamics which is discussed next.
 
\subsection{Determination of the slow Green's function and the criteria for slowness}
The slow Green's function has to be determined from its inverse given in eq. \eqref{eq: inverse slow GF}. Thus, we can determine it from the formal equation,
\begin{equation}
G_s^{-1}\, G_s^{\phantom{-1}} = \mathbf{1}\ .
\end{equation}
Writing the time-space explicitly, we obtain for retarded and advanced part of Keldysh-space,
\begin{equation}
\int dt'\, [G_s^{-1}]^{R/A}(t,t')\, G_s^{R/A}(t',t'') = \delta (t-t'')\ , \label{eq: retarded and advanced kinetic eq}
\end{equation}
and, by use of the gradient expansion, it follows,
\begin{equation}
[G_s^{-1}]^{R/A}(\cmt, \omega)\, e^{-\frac{i}{2} ( \overleftarrow{\partial}_{\cmt} \overrightarrow{\partial}_\omega - \overleftarrow{\partial}_\omega \overrightarrow{\partial}_{\cmt})}\, G_s^{R/A}(\cmt,\omega) = 1\ ,
\end{equation}
where the arrows indicate on which function to apply the derivative. The formal $\tilde{\ \ }\ $-notation is dropped here and for the Keldysh part, for which we obtain,
\begin{equation}
G_s^K(t,t''')\! =\! -\!\! \int\!\!\! dt' \!\!\! \int\!\!\! dt''\, G_s^R(t,t') [G_s^{-1}]^K(t',t'') G_s^A (t'',t''')\ ,
\end{equation}
where $[G_s^{-1}]^K(t',t'') = -(R_k^\dagger \Sigma^K R_k)(t',t'')$.
Application of the gradient expansion yields,
\begin{eqnarray}
& & \hspace{-0.6em} G_s^K(\cmt, \omega)= - \mathrm{exp} \left[{-\frac{i}{2} (\partial_{\cmt}^{\bar K \bar A} \partial_\omega^R + \partial_{\cmt}^{R \bar A} \partial_\omega^K + \partial_{\cmt}^{R K} \partial_\omega^A )}\right] \times \nonumber \\ & & \hspace{5.8em} \times G_s^R(\cmt, \omega) [G_s^{-1}]^K(\cmt, \omega) G_s^A (\cmt, \omega)\ ,
\end{eqnarray}
where in the superscripts of derivatives $R,K,A$ is a compact notation for the corresponding component of the (inverse) Green's function.

Keeping only the zeroth order term of the gradient expansion yields,
\begin{eqnarray}
& &\hspace{0em} [G_{s}^{-1}]^{R/A}(\cmt, \omega)\, G_{s0}^{R/A}(\cmt,\omega) = 1\ , \label{eq: retarded advanced zero}\\
& &\hspace{-4em} G_{s0}^K(\cmt, \omega)= - G_{s0}^R(\cmt, \omega)\, [G_s^{-1}]^K(\cmt, \omega)\, G_{s0}^A (\cmt, \omega)\ ,
\end{eqnarray}
from which we immediately obtain the retarded/advanced Green's function,  eq. \eqref{eq: main slow retarded GF}. In turn, we also obtain the Keldysh Green's function, eq. \eqref{eq: main slow keldysh GF}. In the main text, we dropped the index $0$ for zeroth order.

From the negligibility of the higher order terms, we obtain the criteria for slowness of coordinates. The first order correction to equation \eqref{eq: retarded advanced zero} for the retarded/advanced slow Green's function vanishes, i.e.
\begin{equation}
-\frac{i}{2} [G_s^{-1}]^{R/A}(\cmt, \omega) \, \Big( \overleftarrow{\partial}_{\cmt} \overrightarrow{\partial}_\omega - \overleftarrow{\partial}_\omega \overrightarrow{\partial}_{\cmt}\Big)\, G_{s0}^{R/A}(\cmt,\omega) = 0\ ,
\end{equation}
where we used the zeroth order result for the Green's function. The second order correction reduces to,
\begin{eqnarray}
& &\hspace{-3em} - \frac{1}{8} [G_s^{-1}]^{R/A}(\cmt, \omega) \Big( \overleftarrow{\partial}_{\cmt} \overrightarrow{\partial}_\omega \Big)^2 G_{s0}^{R/A}(\cmt,\omega)=  \nonumber \\ 
& & \hspace{-2em} = \mp \frac{i}{4} \frac{\Gamma''_\sigma(\theta(\cmt))\, \dot \theta^2(\cmt) +\Gamma'_\sigma(\theta(\cmt)) \ddot \theta(\cmt)}{(\omega - \xi_{\alpha \sigma} \pm i \Gamma_\sigma (\theta(\cmt)))^3}\ .
\end{eqnarray}
At resonance $\omega = \xi_{\alpha \sigma} $, this correction is negligible if $\theta(\cmt)$ is slow, such that,
\begin{equation}
1 \gg \frac{\dot \theta(\cmt)}{\Gamma_\sigma(\theta(\cmt))} \approx \mathcal{O} \left(\frac{1}{S}\right)\ , \label{eq: slowness 1}
\end{equation}
where we assumed that $[\partial_\theta \Gamma_\sigma(\theta)]_{\theta= \theta(\cmt)} \approx \mathcal{O} \left( \Gamma_\sigma(\theta) \right)$.

The same criterion for slowness is also relevant for the Keldysh part, but it is not sufficient. For the corrections to the Keldysh part to be negligible, we need two more criteria: First, for the time-derivative acting on the distribution function in $[G_s^{-1}]^K(\cmt, \omega)$, we also need, 
\begin{equation}
1 \gg \frac{\ddot \phi(\cmt)}{T\, \Gamma_\sigma(\theta(\cmt))}\ ; \label{eq: slowness 2}
\end{equation}
Second, for the frequency derivative acting on the distribution function in $[G_s^{-1}]^K(\cmt, \omega)$, we need,
\begin{equation}
1 \gg \frac{ \dot \theta (\cmt)}{T}\ ,\label{eq: slowness 3}
\end{equation}
where $T= \mathrm{min}(T_l , T_r)$.

In conclusion, we have three criteria for slowness from the gradient expansion, eqs. \eqref{eq: slowness 1}, \eqref{eq: slowness 2}, \eqref{eq: slowness 3}. We also have two criteria from the consideration for the slowly rotated self-energy, i.e. both $\theta$ and $\dot \phi$ should be approximately constant on the time-scale $\tau_T= 1/T$. 
These can be summarized in more physical terms: For coordinates to be slow, they should typically change on time-scales much larger than the correlation time of thermal noise and the life-time of electrons on the dot. These conditions are met by $\theta, \dot \phi$ for large spin $S$ (resp. magnetization $M_0$) and not too low temperatures of the leads $T_{l/r}$.

We emphasize, again, a subtle but important point: It is $\dot \phi$ which has to be a slow variable; the angle $\phi$ itself, may change on shorter time-scales. This is important because $\phi$ does change on the time-scale of $1/B$. Thus, it is not necessarily slow. If $\phi$ is slow, then it follows $B \ll \Gamma_\sigma(\theta)$. This would be fatal for the interesting shifts in the distribution function arising from the precession of the magnetization $\sigma\, \omega_\pm = \mathcal O (B)$, since those would be smaller than the level broadening $\sigma\, \omega_\pm \ll \Gamma_\sigma(\theta)$. The interesting case is, thus, for faster precession $B \gg \Gamma_\sigma(\theta)$. Then, $\phi$ is not a slow variable. However, for the approach presented in this article, it is sufficient that $\dot \phi$ is a slow variable.

\bibliography{reference}

\end{document}